\documentclass[aps,amssymb,floatfix,prd,amsmath,preprintnumbers]{revtex4-2}
\usepackage{epstopdf}
\usepackage{capt-of}
\usepackage{graphicx}  
\usepackage{dcolumn}   
\usepackage{bm}
\begin{document}
\input epsf.tex
\title{Cosmological models with Big rip and Pseudo rip Scenarios in extended theory of gravity}
\author{Pratik P Ray \footnote{Department of Mathematics (SAS), Vellore Institute of Technology-Andhra Pradesh University, Andhra Pradesh - 522237, India, E-mail: pratik.chika9876@gmail.com}, Sankarsan Tarai\footnote{Centre of High Energy and Condensed Matter Physics, Department of Physics, Utkal University, Vani Vihar, Bhubaneswar, India-751004 E-mail: tsankarsan87@gmail.com}, B. Mishra \footnote{Department of Mathematics, Birla Institute of Technology and Science-Pilani, Hyderabad Campus, Hyderabad-500078, India, E-mail:bivu@hyderabad.bits-pilani.ac.in}, S.K. Tripathy \footnote{Department of Physics, Indira Gandhi Institute of Technology, Sarang, Dhenkanal, Odisha 759146, India, E-mail:tripathy\_sunil@rediffmail.com}
}
\affiliation{ }

\begin{abstract}
In this paper, we have presented the big rip and pseudo rip cosmological models in an extended theory of gravity. The matter field is considered to be that of perfect fluid. The geometrical parameters are adjusted in such a manner that it matches the prescriptions given by cosmological observations, to be specific to the $H_0$ range. The models favour phantom behaviour. The violation of strong energy conditions are shown in both the models, as it has become essential in an extended gravity. The representative values of the coupling parameter are significant on the evolution of the universe. 
\end{abstract}
\maketitle
\section{Introduction} 
The most serious future dark energy singularity known as the big rip (BR) singularity can occur both in the context of general relativity (GR) and modified or extended theories of gravity. The BR singularity can be characterized as the blowing up of the curvature space- time at finite time. The size of the universe, its expansion and acceleration all diverge \cite{Jambrina06}. It is well informed that the early universe  started from the singular point, known as big bang singularity. We may experience the future universe may enter into the quintessence or phantom phase, that leads to finite time future singularity. The occurrence of future singularity depends on the specific model and the value of effective equation of state (EoS) parameter \cite{Bamba12}. Recently, several cosmological models have been presented on the rip cosmology. Darbowski and coauthors \cite{Darbowski06,Balcerzak06} have tested BR singularity in future state of universe. Nojiri et al. \cite{Nojiri12} have estimated the little rip (LR) dissolution of bound structures in $f(R)$ theory of gravity. Frampton et al. \cite{Frampton12a} have examined the pseudo rip (PR), LR and BR cosmological models and have shown that the PR models can produce an inertial force that does not increase monotonically. However, it decreases after attaining the peak value at a particular future time.\\

Granda and Loaiza \cite{Granda12} have shown that the occurrence BR is based on the parametric value of the model obtained with the kinetic and Gauss-Bonnet coupling. Brevik et al. \cite{Brevik13} have described the phenomena of LR and pseudo phenomena rip in coupled dark energy cosmological models. Makarenko et al. \cite{Makarenko13} have derived a rip model in $f(R,G)$ gravity and have shown the finite time future singularity.
Saez-Gomez \cite{Gomez13} has studied the LR and PR in the modified $f(R)$ gravity. Lopez et al. \cite{Lopez15} have presented the little siblings of the BR which is smoother than the BR singularity. Houndjo et al. \cite{Houndjo14} have investigated the LR cosmology in $f(R,T)$ gravity. In $f(R)$ gravity, Vasilev et al. \cite{Vasilev19} have studied the quantum and classical fate of the BR cosmology.   Mishra and Tripathy \cite{Mishra20} have studied the LR model that evolves in the phantom-like region and at late times overlap with $\Lambda$CDM model. In $f(T)$ gravity, Hanafya and Saridakis \cite{Hanafya20} have shown the ever lasting PR phase of the universe.\\

The accelerated expansion of the universe compelled to modify the GR, as it has certain limitations to justify the expansion. One among the prominent geometrical modification of GR is the $f(R,T)$ gravity\cite{Harko11}. Several aspects of the cosmological and astrophysical issues have been studied in $f(R,T)$ gravity and also it has successfully satisfied some of the observation results in cosmology. Myrzakulov \cite{Myrzakulov12} has studied the geometrical root of $f(R,T)$ gravity and also has shown that this gravity has been successful in describing the accelerated expansion of the universe. Clifton et al. \cite{Clifton12} have given detailed survey on the modified gravity with the cosmological consequences. The anisotropic and dynamical issues in $f(R,T)$ gravity have been discussed in details \cite{Mishra18a,Mishra18b}. Saridakis et al. \cite{Saridakis20} have investigated the cosmological applications of  $f(R,T)$ gravity. Because of the success of $f(R,T)$ gravity in different cosmological aspects, here we are motivated to examine the occurrence of future singularity in finite time in the context of $f(R,T)$ gravity.\\

The late time acceleration of universe further motivates to study the non standard cosmological singularities such as BR singularity, sudden future singularity, finite density singularity and the problem link between energy conditions violation \cite{Perl1999,Tegmark2004}. Apart from BR, all other singularities are weak singularities by their geodesic incompletness nature. BR singularity tested in phantom like universe $(\omega\leq -1)$, as the smaller pressure is allowed to dominate the current evolution and energy density grows proportionally to the scale factor $a(t)$. This motivates us to study the BR and PR cosmological model in an extended theory of gravity. The paper is organised as: In section II, we have described the basic formalism of the model along with the general form of dynamical parameters, whose behaviours to be studied. Two cosmological models based on BR and PR are presented in section III. The analysis and behaviour of the models are given in section IV and the geometric diagnostic analysis and conclusion are given in section V.

\section{Basic Formalism and Dynamical Parameters}
The Einstein-Hilbert action for $f(R,T)$ gravity has been proposed by Harko et al. \cite{Harko11} with the matter Lagrangian considered to be $\mathcal{L}_m=-p$ as, 
\begin{equation} \label{eq:1}
S=\int d^4x\sqrt{-g}\left[\frac{1}{16\pi} f(R,T)+\mathcal{L}_m\right]
\end{equation}
where $p$ be the pressure. $R$ and $T$ respectively denote the Ricci scalar and trace of energy momentum tensor $T_{ij}$. The motivation behind this consideration is that the non-minimal coupling of matter and geometry provide a viable reason for the cosmic acceleration issue. Three functional forms are proposed \cite{Harko11} as follows: (a) $f(R,T)=R+2f(T)$; (b) $f(R,T)=f_1(R)+f_2(T)$; (c)$f(R,T)=f_1(R)+f_2(R).f_3(T)$. In this problem, we have considered the functional,
\begin{equation}\label{eq:2}
f(R,T)=R+2\Lambda_0+2\gamma T
\end{equation}  
where, $\Lambda_0$ is the cosmological constant that does not depend on the cosmic time and $\gamma$ denotes the coupling parameter. So,  $\frac{\partial f(R)}{\partial R} =1$ and $\frac{\partial f(T)}{\partial T}=2\gamma$. Now, the field equitations of $f(R,T)$ gravity can be reduced to the following form,

\begin{equation} \label{eq:3}
R_{ij}-\frac{1}{2}Rg_{ij}= (8\pi+2\gamma) T_{ij} + [\Lambda_0+\gamma\left(\rho-p\right)] g_{ij},
\end{equation}
To frame the cosmological model, we consider an anisotropic Bianchi $VI_h, h=-1$ \cite{Tripathy16} space-time in the form,

\begin{equation}\label{eq:4}
ds^2 = dt^2 - A^2dx^2- B^2e^{2x}dy^2 - C^2e^{-2x}dz^2,
\end{equation}

The metric potentials $A$, $B$ and $C$ are function of cosmic time $t$ only. The matter field is assumed to be that of perfect fluid and the corresponding energy momentum tensor is, $T_{ij}=(\rho+p)u_{ij} - pg_{ij}$. Here, $\rho$ be the matter energy density and $u^{i}u_{i}=-x^{i}x_{i}=1, u^{i}x_{i}=0$. We can derive the $f(R,T)$ gravity field eqn. \eqref{eq:3} in an anistropic space-time with the considered functional of $f(R,T)$ as follow,

\begin{eqnarray} 
\frac{\ddot{B}}{B}+\frac{\ddot{C}}{C}+\frac{\dot{B}\dot{C}}{BC}+ \frac{1}{A^2}&=& -\beta p+\gamma\rho +\Lambda_0, \label{eq:5}  \\
\frac{\ddot{A}}{A}+\frac{\ddot{C}}{C}+\frac{\dot{A}\dot{C}}{AC}- \frac{1}{A^2}&=&-\beta p+\gamma\rho+\Lambda_0, \label{eq:6}  \\
\frac{\ddot{A}}{A}+\frac{\ddot{B}}{B}+\frac{\dot{A}\dot{B}}{AB}- \frac{1}{A^2}&=&-\beta p +\gamma\rho+\Lambda_0, \label{eq:7} \\
\frac{\dot{A}\dot{B}}{AB}+\frac{\dot{B}\dot{C}}{BC}+\frac{\dot{C}\dot{A}}{CA}-\frac{1}{A^2}&=& -\gamma p+ \beta \rho +\Lambda_0, \label{eq:8}  \\
\frac{\dot{B}}{B}&=&\frac{\dot{C}}{C} \label{eq:9}.
\end{eqnarray} 
 
An over dot on the metric potential represents ordinary derivative with respect to the cosmic time. Here $\beta=8\pi+3\gamma$. We can also express the above set of field equations in the form of directional Hubble rate as  $H_x=\frac{\dot{A}}{A},H_y=\frac{\dot{B}}{B},H_z=\frac{\dot{C}}{C}$ and the mean Hubble rate as, $H=\frac{\dot{\mathcal{R}}}{\mathcal{R}}=\frac{1}{3}(H_x+2H_y)$, since eqn. \eqref{eq:9} gives, $H_y=H_z$. The relation between Hubble rate and shear scalar suggests the expression, $H_x=kH_y$, which will enable to have an  anisotropic relationship among the spatial directions. So, eqns. \eqref{eq:5}-\eqref{eq:9} transform to the following equations,
\begin{eqnarray}
6\left(\frac{1}{k+2}\right)\dot{H}+27\left(\frac{1}{k^2+4k+4}\right)H^2+\mathcal{R}^{-\left(\frac{6k}{k+2}\right)}=-\beta p+\gamma\rho +\Lambda_0 \label{eq:10} \\
3\left(\frac{k+1}{k+2}\right)\dot{H}+ 9\left(\frac{k^2+k+1}{k^2+4k+4}\right)H^2-\mathcal{R}^{-\left(\frac{6k}{k+2}\right)}=-\beta p +\gamma\rho+\Lambda_0\label{eq:11}\\
9 \left(\frac{2k+1}{k^2+4k+4}\right)H^2- \mathcal{R}^{-\left(\frac{6k}{k+2}\right)}=-\gamma p+\beta \rho +\Lambda_0 \label{eq:12}
\end{eqnarray}
On solving eqns. \eqref{eq:10}-\eqref{eq:12}, we obtained the matter pressure and energy density in the following form,

\begin{eqnarray} 
p&=&\frac{\gamma}{(\beta^2-\gamma^2)}\left[\frac{3(k^2+k-2)\dot{H}+9(k^2-k-3)H^2}{(k^2+4k+4)}\right] \nonumber \\ 
&-&\frac{\beta}{(\beta^2-\gamma^2)}\left[\frac{3(k^2+3k+2)\dot{H}+9(k^2+k+1)H^2}{(k^2+4k+4)}\right] +\frac{\Lambda_0}{(\beta+\gamma)} \label{eq:13}\\
\rho&=&\frac{\beta}{(\beta^2-\gamma^2)}\left[\frac{9(2k+1)}{(k^2+4k+4)}H^2\right] \nonumber \\ 
&-&\frac{\gamma}{(\beta^2-\gamma^2)}\left[\frac{6(k+2)\dot{H}+27H^2}{(k^2+4k+4)}\right]-\frac{\Lambda_0}{(\beta+\gamma)} \label{eq:14}
\end{eqnarray}

We derive the equation of state (EoS) parameter $\omega=\frac{p}{\rho}$ and effective cosmological constant $\Lambda$. The EoS parameter will enable us to make some investigation on the late time cosmic acceleration problem.
\begin{eqnarray}
\omega &=& -1 +(\beta+\gamma)\nonumber \\ 
&\times& \left[\frac{3(k^2+3k+2)\dot{H}+9(k^2-k)H^2}{\gamma\left[6(k+2)\dot{H}+27H^2\right]-\beta\left[9(2k+1)H^2\right]+(k+2)^2(\beta-\gamma)\Lambda_0}\right]\label{eq:15}\\
\Lambda &=&\frac{\gamma}{(\beta+\gamma)}\left[\frac{6\dot{H}+18H^2}{(k+2)}-2\Lambda_0\right]+\Lambda_0 \label{eq:16}
\end{eqnarray}

We have seen that all the dynamical parameters are expressed in terms of the Hubble parameter. In order to frame the cosmological model, we are intending to study the problem with an assumed Hubble parameter, so here we consider the BR and PR cosmological model in the subsequent section.
\section{Cosmological Models with Rip Cosmology}

\subsection{BR Model}
\begin{figure}[!htp]
\centering
\includegraphics[scale=0.50]{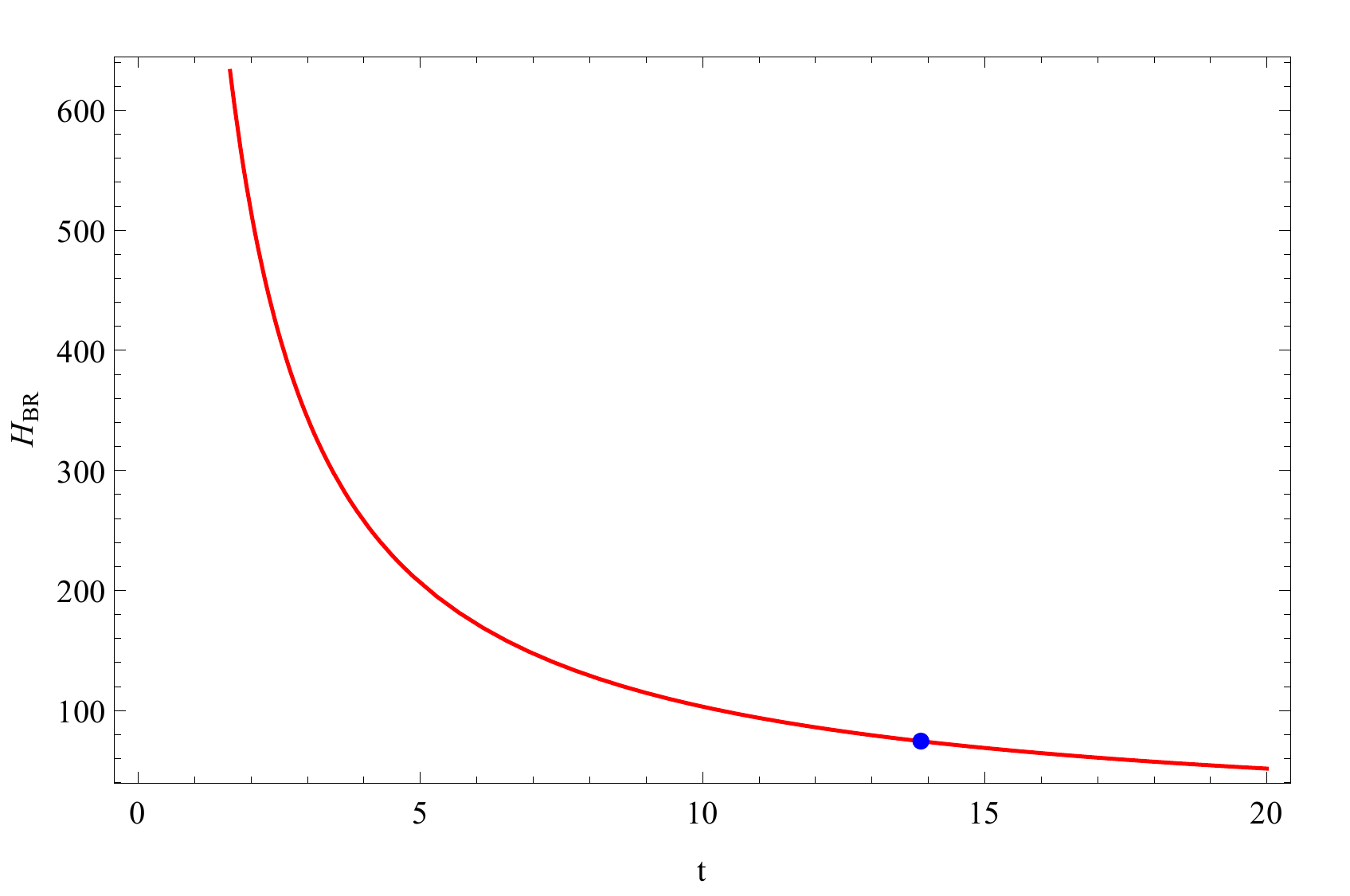}
\caption{Graphical behaviour of Hubble parameter in cosmic time for BR model.}
\label{Fig0}
\end{figure}

It is well known that the cosmological models in the context of extended theory of gravity addressed some issues of accelerating universe. Theoretically, the claim is based on the corresponding EoS parameter, $\omega<-1$, the model which crosses the phantom barrier. Caldwell et al. \cite{Caldwell03} have shown that the phantom cosmological models contain the BR singularity i.e. at the finite time the scale factor diverges. The occurrence of BR in the model dissolute the bounded system \cite{Caldwell03,Frampton03,Newweris04}. The background motivation to BR singularity is taken from phantom scenario  \cite{Caldwell04}. phantom violates the null energy condition (NEC), $\rho + p = - \dot{\phi}^{2}<0,$ where $\phi$ is a scalar field of negative kinetic energy that simulates the phantom scenario. This value of NEC is obtained as the scalar field $\phi$ which dominates the current evolutionary era and executes energy density $\rho = - \frac{\dot{\phi}^{2}}{2} + V(\phi)$ and the pressure $ p = -\frac{\dot{\phi}^{2}}{2} - V(\phi).$  Phantom scenario has been  successfully investigated in many popular cosmological models such as the super string model, brane model, viscous model, Brans-Dicke theory under the framework of GR, where the EoS parameter $\omega < - \frac{3}{2}$. In addition, the energy density of phantom is proportional to the scale factor. So, the growth of energy density corresponds to the increase of scale in order to contribute to the expansion of the universe. This scenario favours a unique type of singularity in the universe that comes to action despite of violation of all energy conditions, known as BR singularities. One can consider it as a perfect singularity in the sense of geodesic incompleteness excluding few isotropic geodesics which are complete. In this case, we consider the BR scale factor in the form,

\begin{equation}
a(t)= \dfrac{1}{(t_{s}-t)^{\alpha}}+ a_{0}, \label{eq:17}
\end{equation}

where $a_{0}$ is the integration constant. The scale factor, $a \rightarrow \infty$ as $t \rightarrow t_{s}.$ Similarly, as $t \rightarrow \infty,$ the scale factor approaches to asymptotic emptiness. Hence, this singularity emerges for phantom-like EoS parameter, $\omega < -1$ \cite{Caldwell03}. The Hubble parameter obtained as, $ H=\frac{\alpha}{t_{s}-t} $, where $t_{s}$ is the moment when BR takes place and at $t=t_s$, the cosmic derivative and Hubble rate blow up. So, at this space time point, the curvature is ill-defined. Here, $ \alpha $ and $t_{s}$ are free model parameters which need to be specified from some physical basis. It is note that, $\dot{H}>0$ whether we consider a phantom-like phase $(t < t_{s})$ or non-phantom-like phase $(t > t_{s})$. The deceleration parameter becomes, $q= -1-\dfrac{1}{\alpha}.$ The negative value of deceleration parameter depends only on the value of the parameter $\alpha$ which decides bouncing behaviour of the model. For positive values of $\alpha,$ the deceleration parameter becomes negative and asymptotically approaches to $-1$ and when $\alpha>1$, it indicates the accelerating behaviour of the universe. Next, we need to fix the unknown model parameters from some physical constraints. By fixing the present value of deceleration parameter as $-1.08$ from a recent analysis \cite{Camarena20}, we obtained the value of $\alpha= 12.7.$ Substituting the value of $\alpha,$ $H= H_{0}\simeq 74.33$ and $t = t_{0}\simeq 13.82$ Gyr. The same has been represented graphically in Fig. \ref{Fig0} and with this now we can constraint the value of the constant parameter $t_{s}$. \\

The dynamical behaviour of the model can be investigated through the determination of matter pressure, energy density and the EoS parameter. The pressure ($p_{BR})$ and  energy density ($\rho_{BR}$) for the BR model can be derived as, 
\begin{eqnarray} 
p_{BR}&=&\frac{\gamma}{(\beta^2-\gamma^2)}\left[\frac{3(k^2+k-2)\alpha+9(k^2-k-3)\alpha^2}{(t_s-t)^2(k^2+4k+4)}\right] \nonumber \\ 
&-&\frac{\beta}{(\beta^2-\gamma^2)}\left[\frac{3(k^2+3k+2)\alpha+9(k^2+k+1)\alpha^2}{(t_s-t)^2)(k^2+4k+4)}\right] +\frac{\Lambda_0}{(\beta+\gamma)} \label{eq:18}\\
\rho_{BR}&=&\frac{\beta}{(\beta^2-\gamma^2)}\left[\frac{9(2k+1)\alpha^2}{(k^2+4k+4)(t_s-t)^2)}\right] \nonumber \\ 
&-&\frac{\gamma}{(\beta^2-\gamma^2)}\left[\frac{6(k+2)\alpha+27\alpha^2}{(k^2+4k+4)(t_s-t)^2}\right]-\frac{\Lambda_0}{(\beta+\gamma)} \label{eq:19}
\end{eqnarray}

\begin{figure}[!htp]
\centering
\includegraphics[scale=0.50]{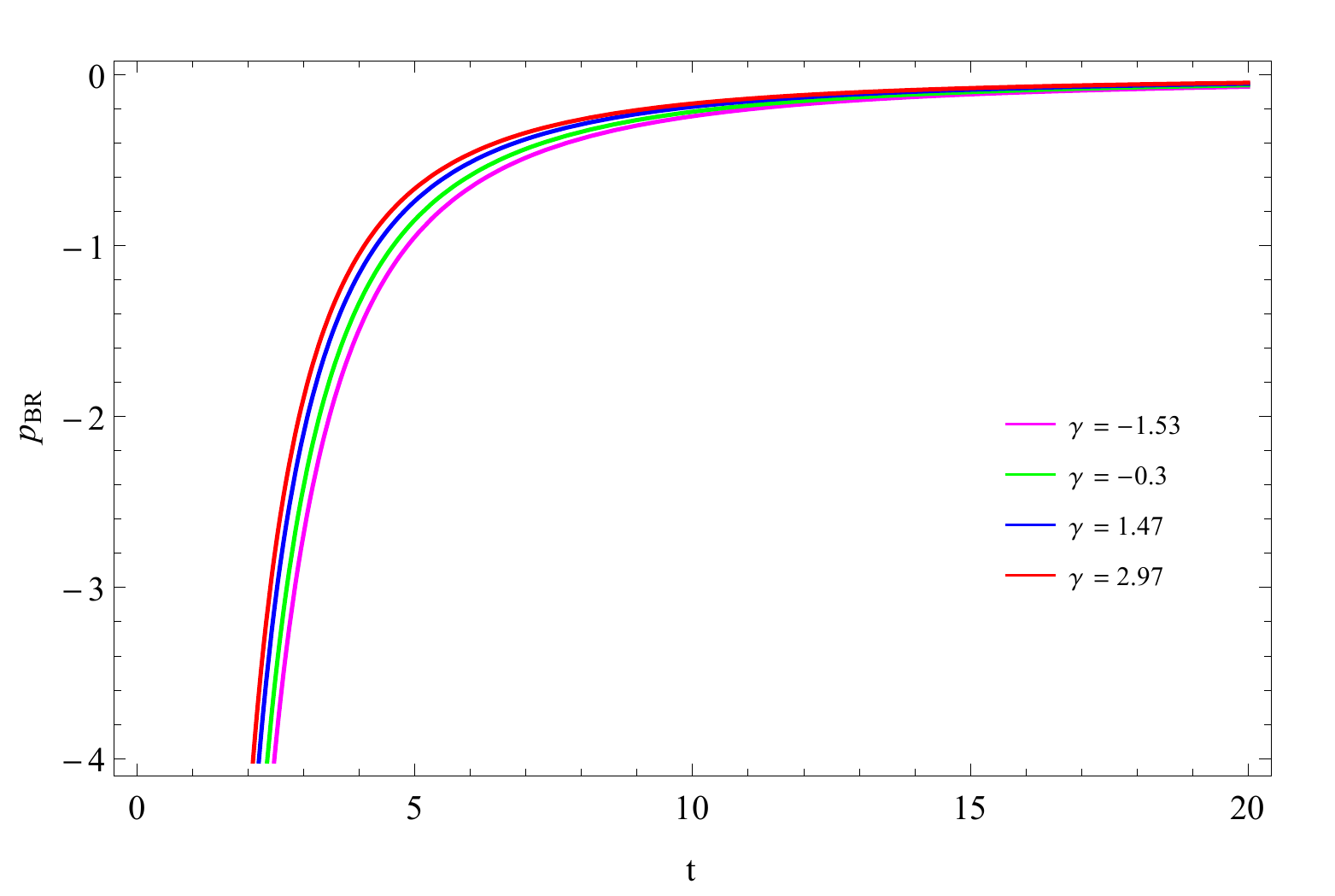}
\includegraphics[scale=0.50]{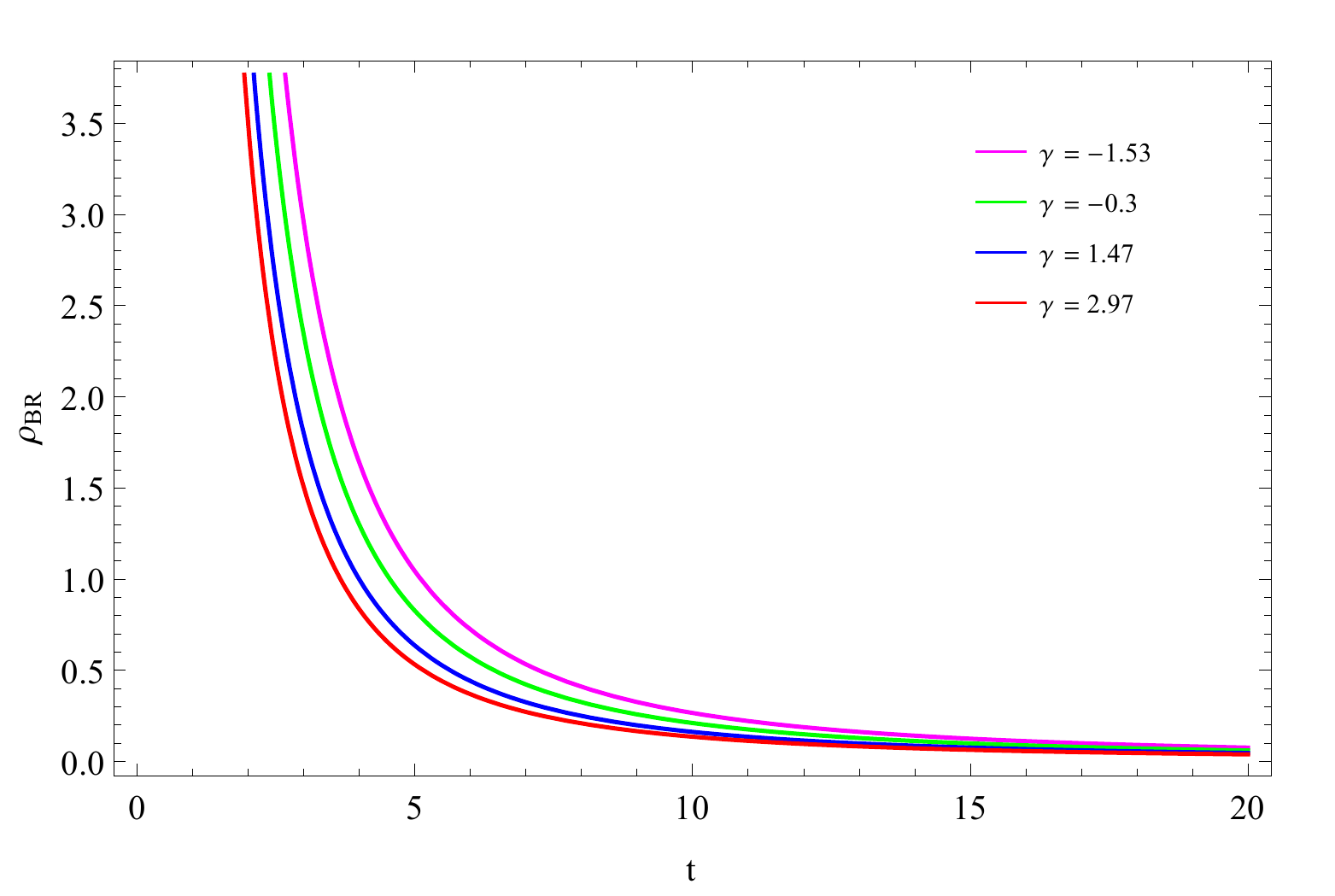}
\caption{Graphical behaviour of pressure (left panel) and energy density (right panel) in cosmic time of BR model with the representative value of the parameters, $k=1.0000814$, $\alpha=12.5$, $t_s=0.1$, $\Lambda_0=-0.25$.}
\label{Fig1}
\end{figure}

The evolution of pressure and energy density of the model depends on model parameters; $\gamma,$ $k,$ $\alpha,$ $t_{s}$ and $\Lambda_{0}.$ We consider the value of the anisotropy parameter $k$ and coupling parameter $\gamma$ to maintain the negative pressure and positive energy density through out the cosmic evolution. However, we can change the values of $\gamma$ and $t_{s}$ in order to analyse the behaviour of the dynamical parameters. It is worth to mention that the energy density and pressure diverges with a finite time, i.e, as $t \rightarrow t_{s},$ $\rho_{BR} \rightarrow \infty$ and $\mid p_{BR} \mid \rightarrow \infty.$ So, $t_{s}$ corresponds to the life time of the universe. Moreover, we can not demonstrate the behaviour of the model in the absence of coupling parameter. Because, as $\gamma \rightarrow 0,$ the energy density and pressure become independent of time and approaches to constant values $\dfrac{\Lambda_{0}}{\beta + \gamma}$ and $- \dfrac{\Lambda_{0}}{\beta + \gamma}$ respectively. Since the value of $\Lambda_{0}$ is fixed to be $-0.25$, then energy density becomes negative and pressure becomes positive for $\gamma=0,$ which is unphysical. In addition, the role of non-zero coupling constant is justified with as in Fig. \ref{Fig1}.\\

In Fig. \ref{Fig1} (left panel), the matter pressure arises with a higher negative value at early epoch and increases monotonically to a small negative value $\thickapprox 0$, as $t \rightarrow t_{s}$. In Fig. \ref{Fig1} (right panel), the energy density starts evolving at different values of $\gamma$ and gradually decreases to reach lower positive values as $t \rightarrow t_{s}$. As the value of the coupling parameter increases, the growth rate of energy density shows decreasing behaviour, however the growth rate of matter pressure of the model is proportional to the values of coupling constants in the negative domain. Another important observation is that both the pressure and energy density can merge as $t \rightarrow t_{s}$ and remain constant throughout to indicate $\rho_{BR} \rightarrow \infty$ and $\mid p_{BR} \mid \rightarrow \infty$. We remark that, the pressure and energy density in the BR model depends vitally on the value of coupling parameter. On the other hand from eqns. \eqref{eq:18}-\eqref{eq:19}, the EoS parameter $\omega_{BR}=\frac{p_{BR}}{\rho_{BR}}$, which will enable us to study the late time acceleration issue and the effective cosmological constant $\Lambda_{BR}$ can be obtained as.
\begin{eqnarray}
\omega_{BR} &=& -1 +(\beta+\gamma)\nonumber \\ 
&\times& \left[\frac{3(k^2+3k+2)\alpha+9(k^2-k)\alpha^2}{\gamma\left[6(k+2)\alpha+27\alpha^2\right]-\beta\left[9(2k+1)\alpha^2\right]+(k+2)^2(\beta-\gamma)\Lambda_0}\right]\label{eq:20}\\
\Lambda_{BR} &=&\frac{\gamma}{(\beta+\gamma)}\left[\frac{6\alpha+18\alpha^2}{(k+2)(t_s-t)^2}-2\Lambda_0\right]+\Lambda_0 \label{eq:21}
\end{eqnarray}

\begin{figure}[!htp]
\centering
\includegraphics[scale=0.50]{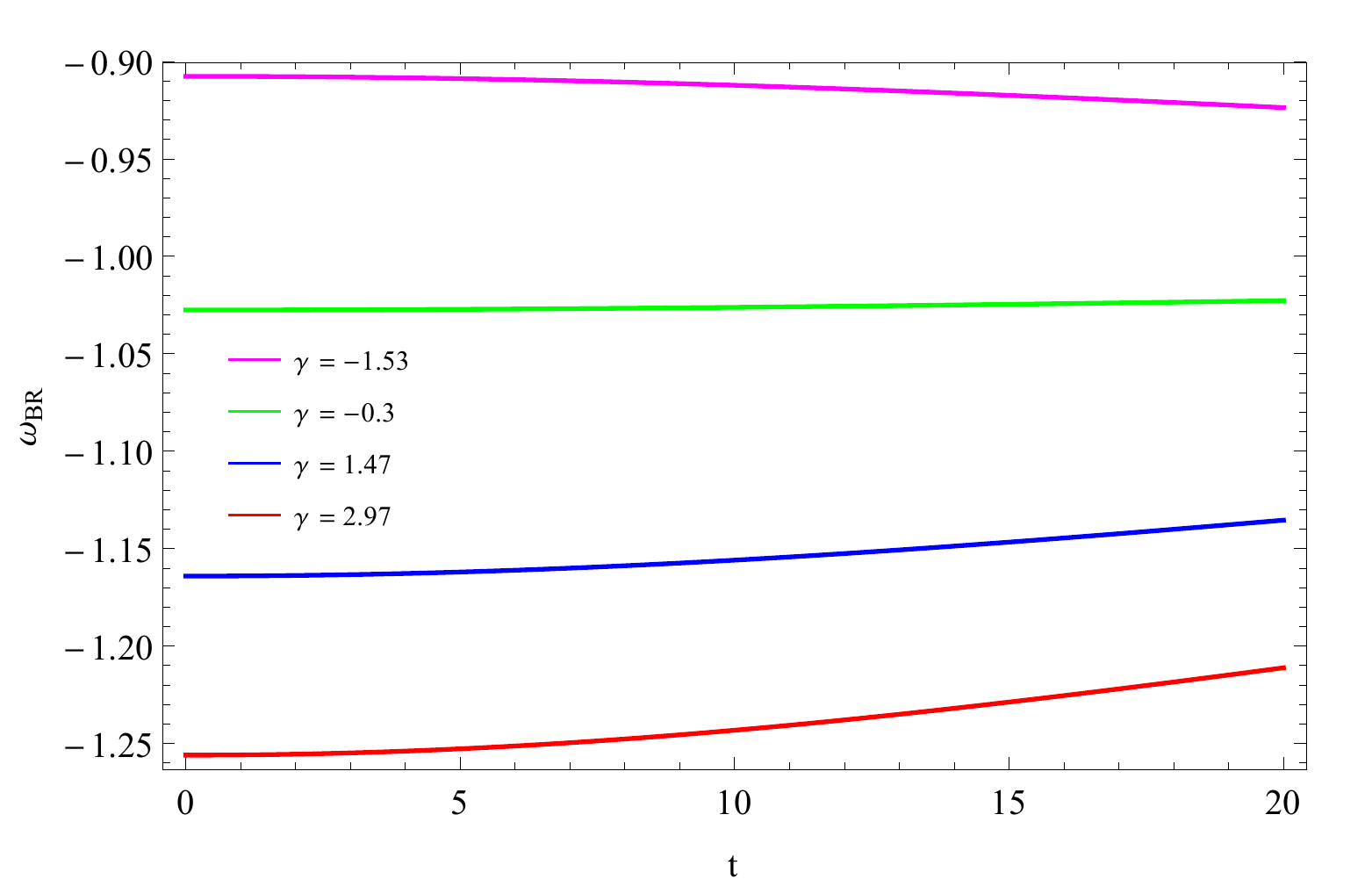}
\includegraphics[scale=0.50]{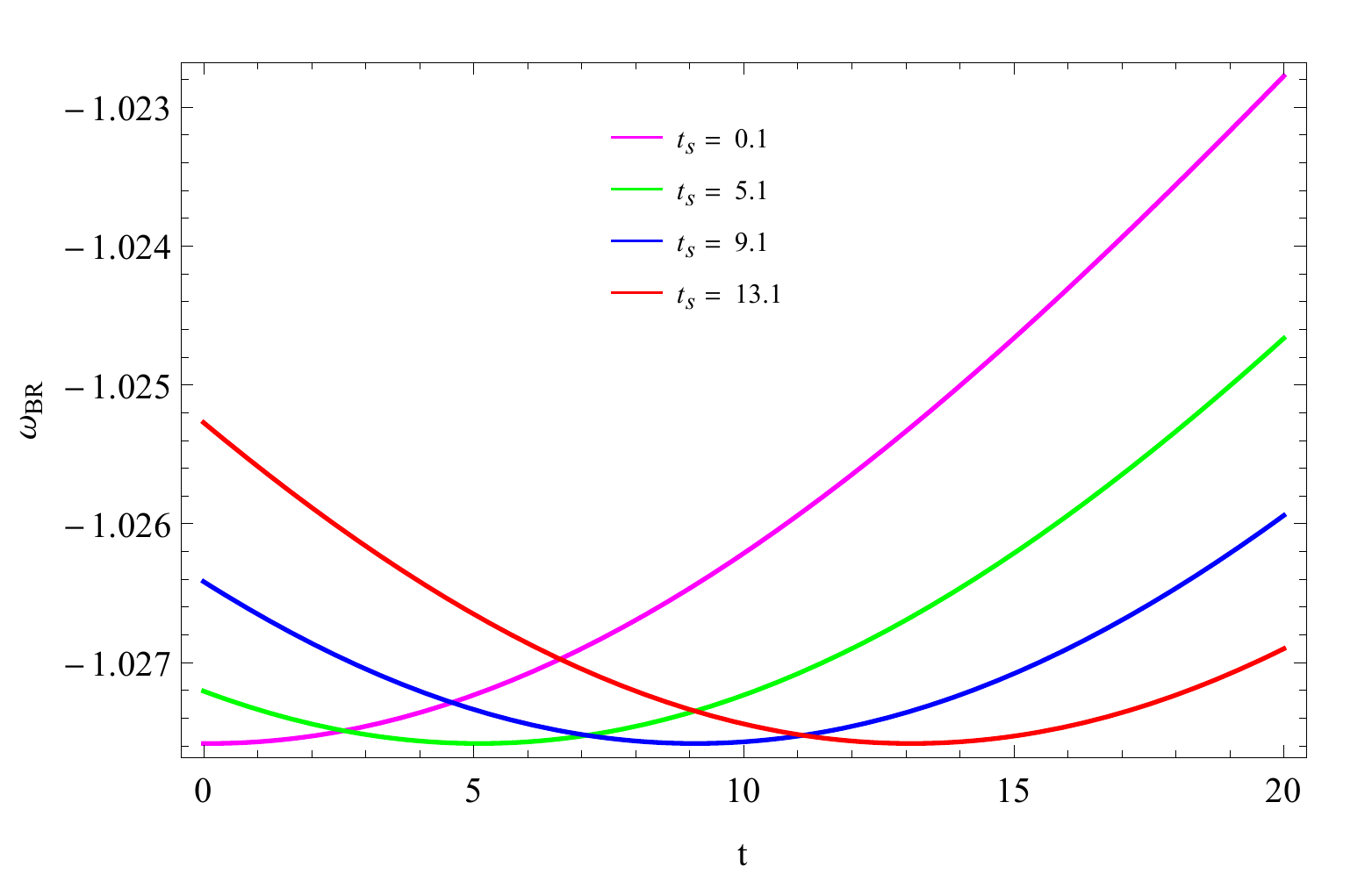}
\caption{Graphical behaviour of EoS parameter in cosmic time with the representative values of $\gamma$, $t_s=0.1$ (left panel); with the representative values of $t_s$, $\gamma=-0.3$ (right panel) for BR model. The other parametric values are $k=1.0000814$, $\alpha=12.5$, $\Lambda_0=-0.25$.}
\label{Fig2}
\end{figure}

The dependency of the EoS parameter on the model parameters has been inevitable. If $\alpha \rightarrow 0$, the present model may favours the $\Lambda$CDM behaviour because $\omega_{BR} \rightarrow -1$. The graphical behaviour of $\omega_{BR}$ and $\Lambda_{BR}$ given in Fig. \ref{Fig2}, where the analysis on the role of $\gamma$ and $t_{s}$ are explained.

In Fig. \ref{Fig2} (left panel), the curve of EoS parameter starts evolving in the phantom phase at the initial epoch, increases gradually and remains in the same region approaching to the $\Lambda$CDM line towards the late phase of evolution. At present cosmic epoch, BR model records the EoS parameter to be $\omega_{BR}= -1.025$ (Green line) which is in nice agreement with the range $\omega(t_{0}) = -1.10 \pm 0.14$ (68 $\%$ CL) as constraint by a recent observation \cite{Komatsu11}. The evolution of EoS parameter has been analysed for four representative values of coupling parameter $\gamma.$ Also, it is observed that with increasing values of $\gamma$ from negative to positive, the EoS parameter starts shifting from quintessence region and finally restores back in the phantom region. In view this, it is worth to mention here that Nojiri \cite{Nojiri05} have discussed the structures of future singularities including BR within finite time $(t_{s})$ and investigated the fate of phantom driven universe. After examining the BR model of transition on the basis of EoS parameter, Nojiri \cite{Nojiri05} concluded that EoS parameter needs to be doubled valued in order to get a continuous transition from quintessence to phantom phase. Such scenario can be experienced in this model, when there is a first order phase transition. Interestingly, the pink curve of EoS parameter $(\gamma = -1.53)$ that lies in quintessence region, then decreases smoothly towards $\Lambda$CDM line as $t \rightarrow t_{s}$; whereas the green curve, blue curve and red curve show an smooth increasing trend towards $\Lambda$CDM line in the same cosmic time scale. So, the coupling parameter has a significant role on evolution of EoS parameter of BR model and confirms the physical viability of the model as well.\\

\begin{figure}[!htp]
\centering
\includegraphics[scale=0.50]{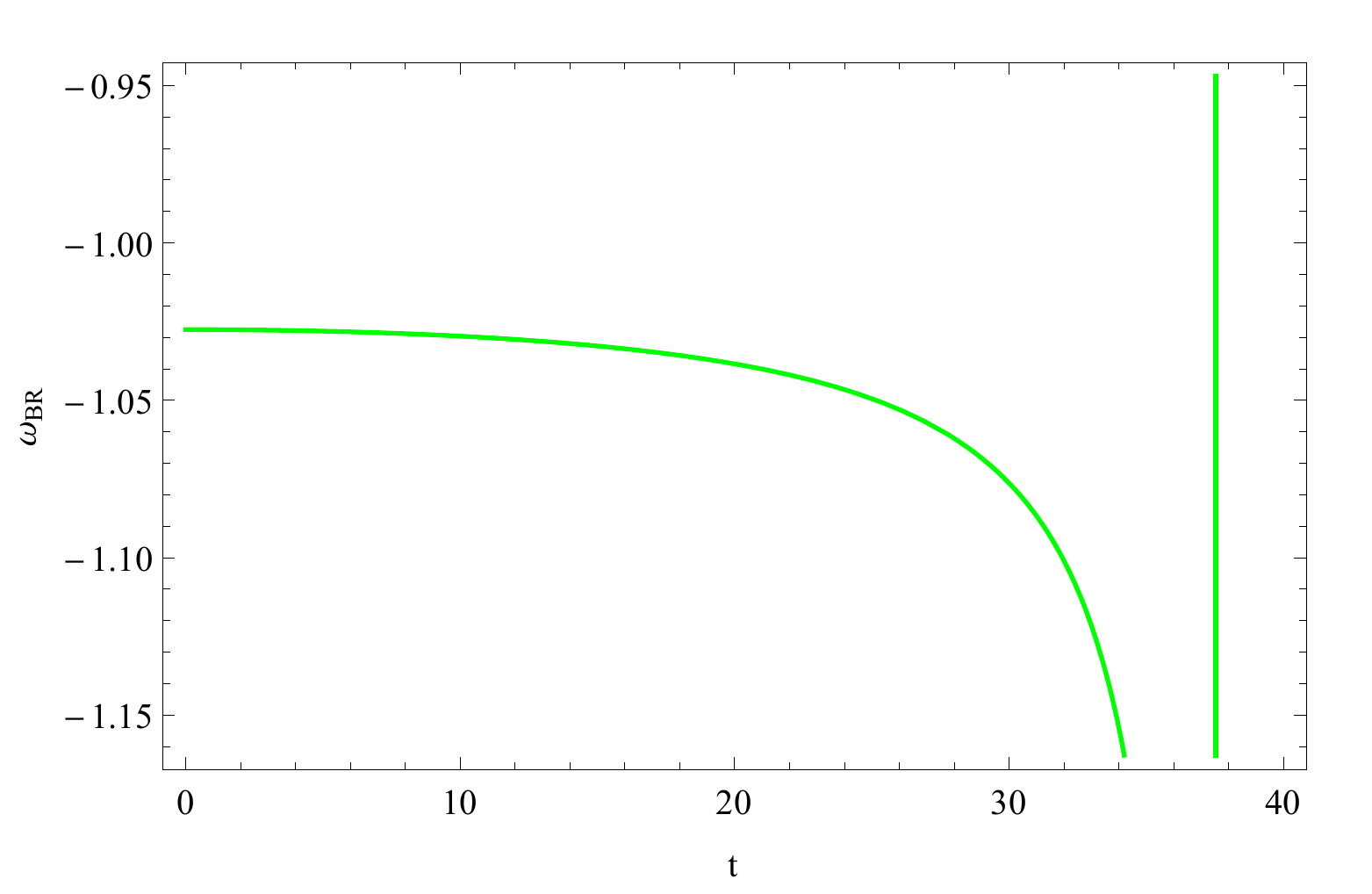}
\includegraphics[scale=0.50]{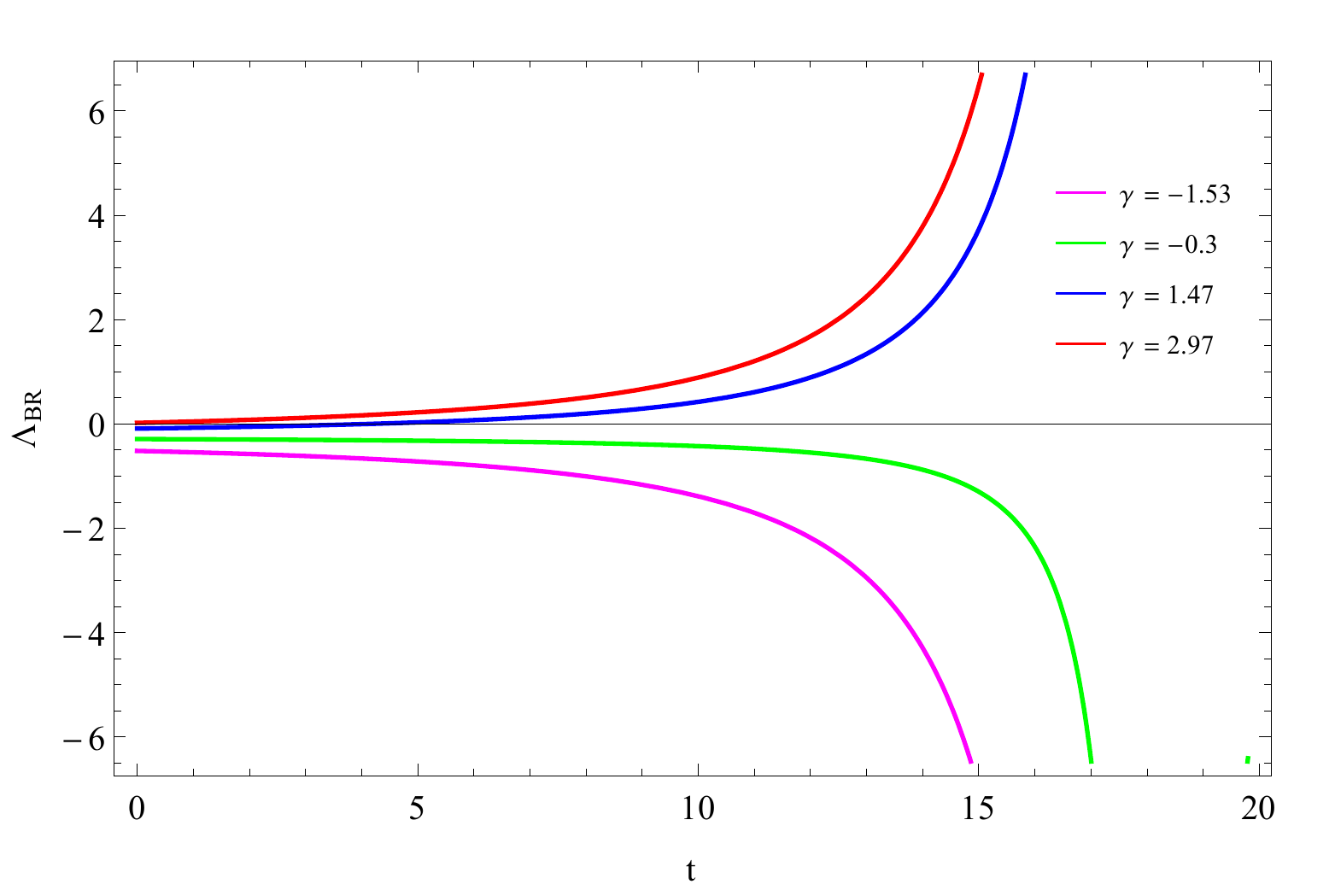}
\caption{Graphical behaviour of EoS parameter showing singularity (left panel), effective cosmological constant (right panel) with the representative values of $\gamma$ in cosmic time for BR model. The other parametric values are $k=1.0000814$, $\alpha=12.5$, $\Lambda_0=-0.25$.}
\label{Fig3}
\end{figure}

Fig. \ref{Fig2} (right panel) demonstrates the dynamics of EoS parameter in cosmic time with representative moments $(t_{s})$ when BR takes place. As the value of $t_{s}$ starts increasing, the curves (from pink to red) start evolving from a lower negative value to higher ones in the phantom region during early epoch. However, as time increases, the curves change their nature surprisingly and remain in the phantom phase in a decreasing order starting from higher negative values (pink to red). The reason behind this sudden change may be due to the dominant nature of phantom in the cosmic evolution era. This behaviour supports that the  BR model is driven by phantom. This confirmation becomes more concrete as a suitable singularity can be achieved by the EoS parameter of present model for specific choice of values of model parameters $t_{s}$ and $\Lambda_{0}$. So, it can be inferred that, the value of cosmological constant also affects the EoS parameter of the model. We can infer from Fig. \ref{Fig3} (left panel) that the BR phase begins at a cosmic time, $t_{0}= 35.1$ Gyr which indicates that the BR scenario would have occurred approximately $22$ billion years before the present cosmic time and matches with the prediction by Caldwell et al.\cite{Caldwell03}. They derived the time duration from present cosmic time till the BR occurrence time as, $t_{s} - t_{0}\approx \dfrac{2}{3 \mid 1+ \omega_{BR} \mid H_{0} \sqrt{1- \Omega_{m}}} ,$ where $\Omega_{m}$ is the value of density of all possible matters in the universe.\\

In Fig. \ref{Fig3} (right panel), the behaviour of effective cosmological constant has been shown. For the negative value of the coupling constant, $\Lambda_{BR}$ turns out to be negative (pink curve, green curve) and decreases rapidly over the time. At the same time, for positive values of coupling constant, $\Lambda_{BR}$ starts evolving as a positive quantity (blue curve, red curve) and grows more rapidly. In both case, the curves of $\Lambda_{BR}$ diverges for $t > 14.95$ Gyr in the late time evolutionary era. It mimics the future singularity behaviour of the present model contributing towards the accelerated expansion of the universe.

\subsection{PR Model}
There is a set of models where the dark energy density increases monotonically with the scale factor, but bounded from above by the limiting density $\rho_{PR}\rightarrow\infty$. This kind of model leads to dissolute the bound structure and termed as PR model \cite{Frampton12}. In this section, we consider the PR scale factor with the corresponding Hubble parameter as, $H=H_0-H_1e^{-\lambda t},$ where $H_0>0$ is the Hubble tension at present cosmic time, $H_{1}>0$ and $\lambda>0$ are constant parameters. The scale factor, thus, can be derived as,
\begin{equation}
a = a_{0}~exp \left[H_{0}t + \frac{H_{1}}{\lambda}e^{- \lambda t}\right] \label{22}
\end{equation}
It can be noted that, as $ t \rightarrow \infty $, the Hubble parameter increases with time asymptotically tending to observational range of Hubble parameter value at present epoch [$74.3 \pm 1.42$ $km^{-1}$ $Mpc^{-1}$] \cite{Aghanim20}. However, the higher derivatives of Hubble parameter diverges for this choice. Although the Hubble parameter remains finite as time goes to infinite but the inertial force dissociates few bound structures of the model through out the process of cosmic evolution. In fact, effect of Hubble expansion on local cosmological models have been explored widely \cite{Gomez13} giving opportunity to many open problems. Also, when $t \rightarrow  \infty $, the scale factor turns out to be de Sitter solution.  Brevik \cite{Brevik12} considered a PR model with asymptotically de Sitter solution in order to investigate on the coupling between dark energy and dark matter. \\

\begin{figure}[h!]
\centering
\includegraphics[scale=0.45]{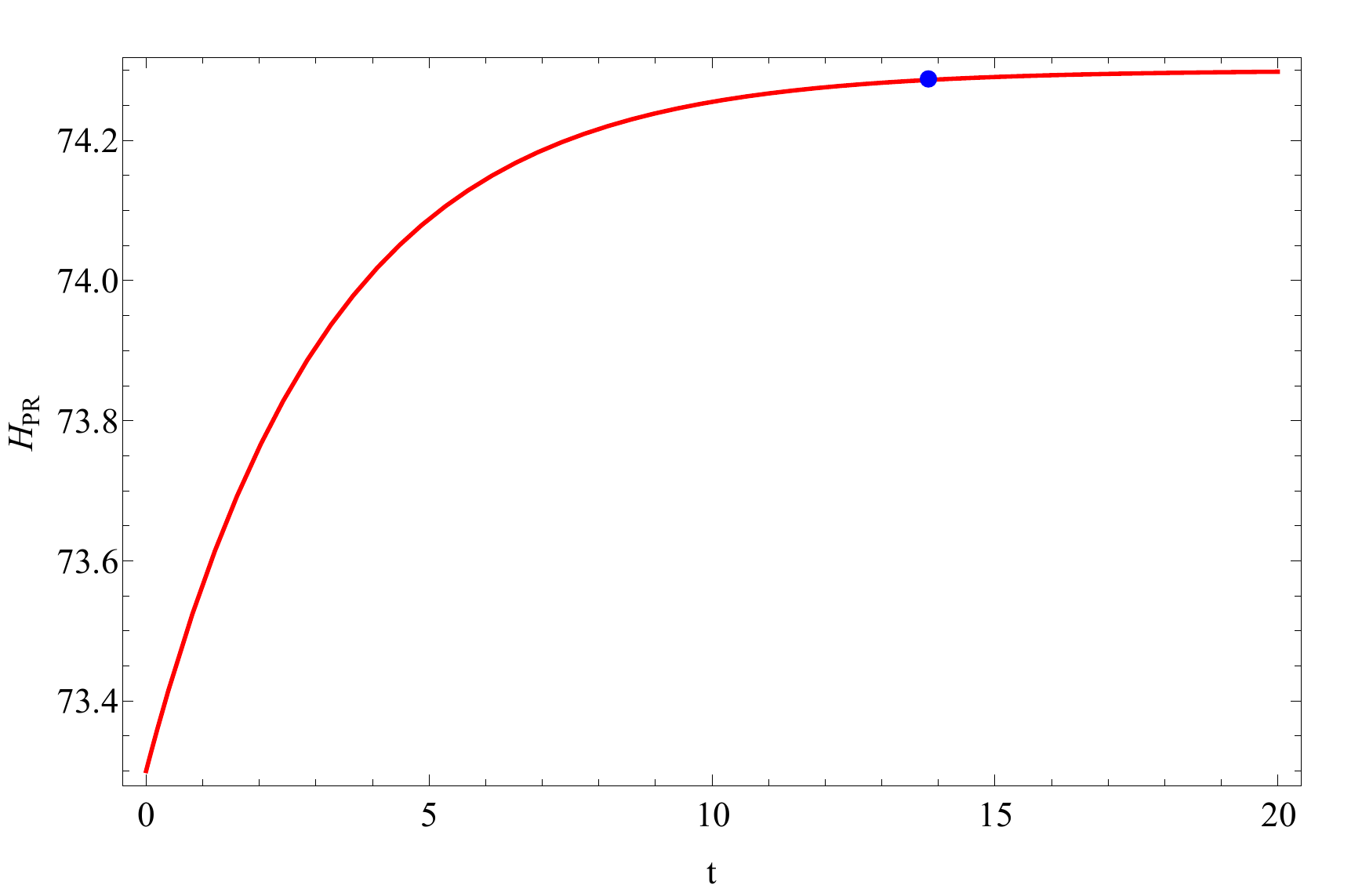}
\includegraphics[scale=0.45]{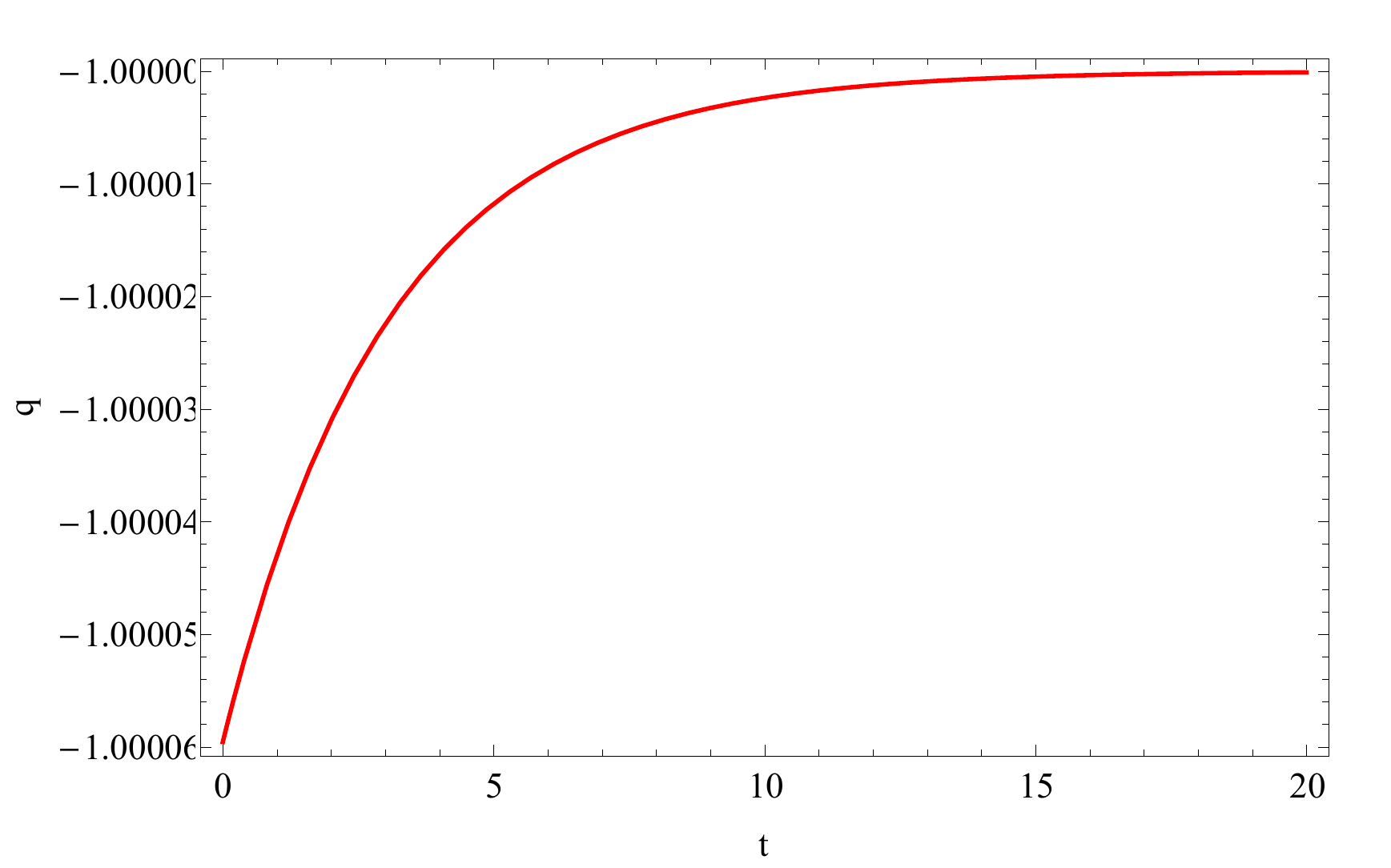}
\caption{ Graphical behaviour of Hubble parameter (left panel) and deceleration parameter (right panel) in cosmic time for the PR model. }  \label{Fig4}
\end{figure}

The Hubble parameter increases with increase in time and the present value of  found to be $H=74.31$ [Fig. \ref{Fig4}, left panel]. The deceleration parameter can be obtained as, $q=-1-\frac{\lambda H_1  e^{-\lambda t}}{(H_0-H_1e^{-\lambda t})^2}$. At $t \rightarrow 0$, $q = -1 - \dfrac{\lambda H_{1}}{H_{0}^{2}}$ and when $t\rightarrow \infty$, $q$ approaches to $-1$. We constrained the parameters $H_{1} > 0$ and $\lambda > 0$ in order to fix the present value of the deceleration parameter  $q_{0}$ in the preferred range of recent observation ($q_{0}= -1.08 \pm 0.29$) \cite{Camarena20}. In, Fig. \ref{Fig4} (right panel), the deceleration parameter increases from a small negative value and approaches to $-1$ at late epoch. For $\lambda > 0,$ $e^{- \lambda t} > 0$, the deceleration parameter remains negative throughout the evolution indicating an ever accelerating behaviour. To mention here, it is not possible to show the transit behaviour from the decelerated to accelerated phase to avoid singularity at infinite time scale. For brevity, we consider $1$ unit of cosmic time as $1 Gyr$. \\

Previously, we obtained, $H_{0}= 74.31$ and for brevity we consider, $H_{1}= 1$ $kms^{-1}$ $ Mpc^{-1}$. This value of present Hubble tension is available in a recent observational analysis \cite{Kenworthy19} which combines data from the Pantheon sample with the foundation survey and the most recent discovery of light curves from the Carnegie Supernova Project. Substituting these above mentioned values in the Hubble formula of PR model, we obtained the value of Hubble parameter at present time to be $H_0 \simeq 74.32$ for $\lambda t = 4.161 $ and lies within the preferred range $(H_0 = 74.3 \pm 1.42)$ as constrained by a recent observation of distance ladder measurement \cite{Riess18}. As a result of the substitution, along with the value of present time $t = t_{0}= 13.82$ Gyr, we obtained the parameter value $\lambda \simeq 0.3011$ $(Gyr)^{-1}$. Similarly, incorporating the value of $H_{0},$ $H_{1}$ and $\lambda,$ we obtained the value of deceleration parameter at present time as $q_{0}= -1.000002$ which lies in the constrained range, $q_{0} = -1.08 \pm 0.29$ according to the latest reviewed results of Plank collaboration \cite{Aghanim20}.\\

We shall investigate the pressure and energy density of the PR model to analyse the evolutionary behaviour. Using the Hubble parameter of PR model, \eqref{eq:13} and \eqref{eq:14} can be expressed as,   
\begin{eqnarray} \nonumber
p_{PR}=&-&\frac{\gamma}{(\beta^2-\gamma^2)}\left[\frac{3(k^2+k-2)\lambda H_1e^{-\lambda t}+9(k^2-k-3)(H_0-H_1e^{-\lambda t})^2}{(k^2+4k+4)}\right] \nonumber \\ 
&-&\frac{\beta}{(\beta^2-\gamma^2)}\left[\frac{3(k^2+3k+2)\lambda H_1e^{-\lambda t}+9(k^2+k+1)(H_0-H_1e^{-\lambda t})^2}{(k^2+4k+4)}\right]\nonumber \nonumber \\
&+&\frac{\Lambda_0}{(\beta+\gamma)}\label{eq:23}\\ 
\rho_{PR} =&+&\frac{\beta}{(\beta^2-\gamma^2)}\left[\frac{9(2k+1)(H_0-H_1e^{-\lambda t})^2}{(k^2+4k+4)}\right] \nonumber \\ 
&-&\frac{\gamma}{(\beta^2-\gamma^2)}\left[\frac{6(k+2)\lambda H_1e^{-\lambda t}+27(H_0-H_1e^{-\lambda t})^2}{k^2+4k+4)}\right]\nonumber \\
&-&\frac{\Lambda_0}{(\beta+\gamma)}\label{eq:24}
\end{eqnarray}

The model parameters $\gamma,$ $H_{1},$ $\lambda,$ $\Lambda_{0}$ control the evolutionary aspects of the dynamical parameters. The anisotropic parameter $k$ has been chosen with the value $1.0000814$ \cite{Mishra18} which results to the value of average anisotropy $\mathcal{A} \approx 4.91 \times 10^{-10},$ where $\mathcal{A} = \dfrac{1}{3} \left(1-\dfrac{H_{i}}{H}\right)^2$ for $i=1,2, 3$ \cite{Jaffe06,Saadeh16}. \\

\begin{figure}[!htp]
\centering
\includegraphics[scale=0.50]{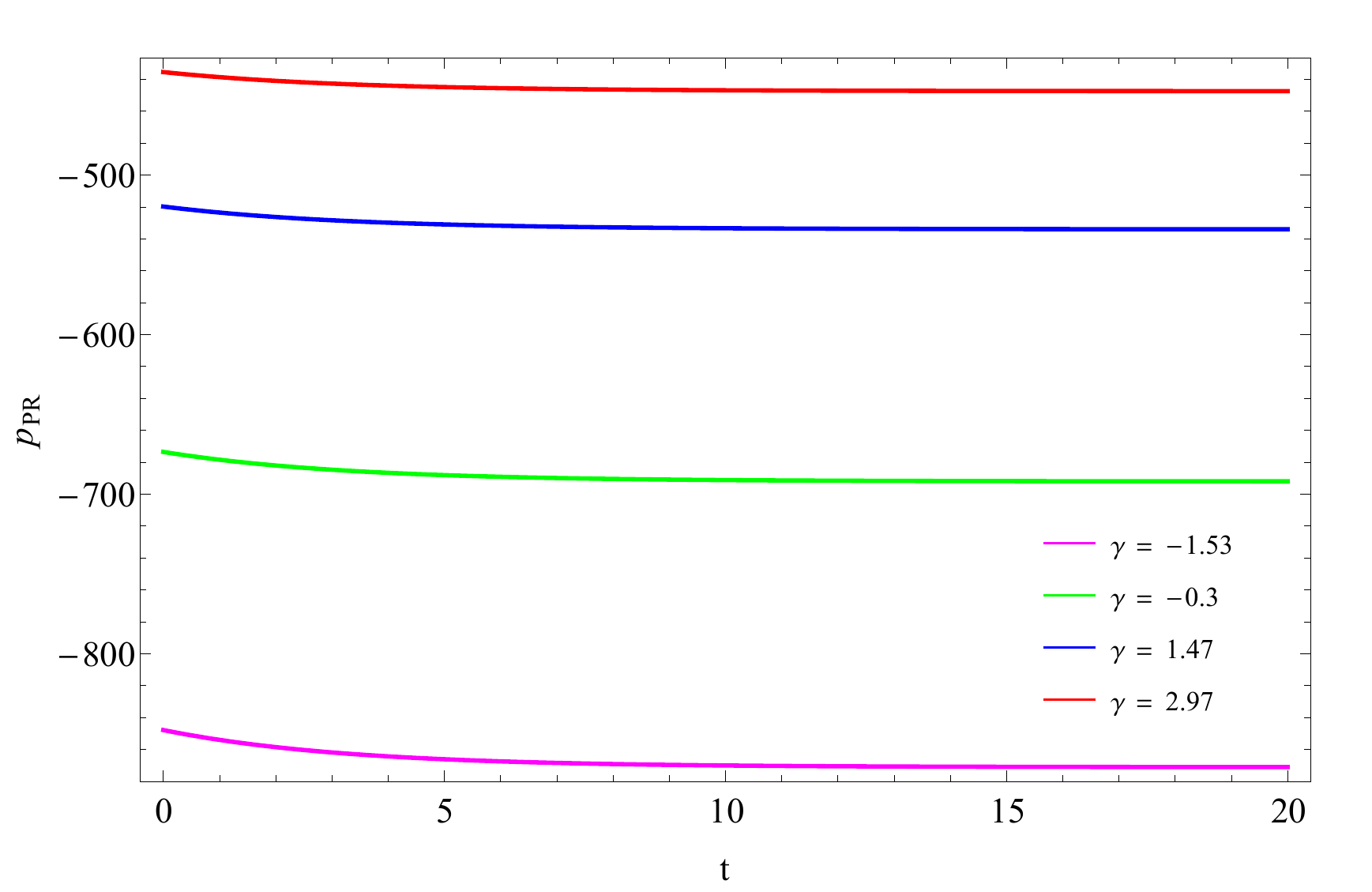}
\includegraphics[scale=0.50]{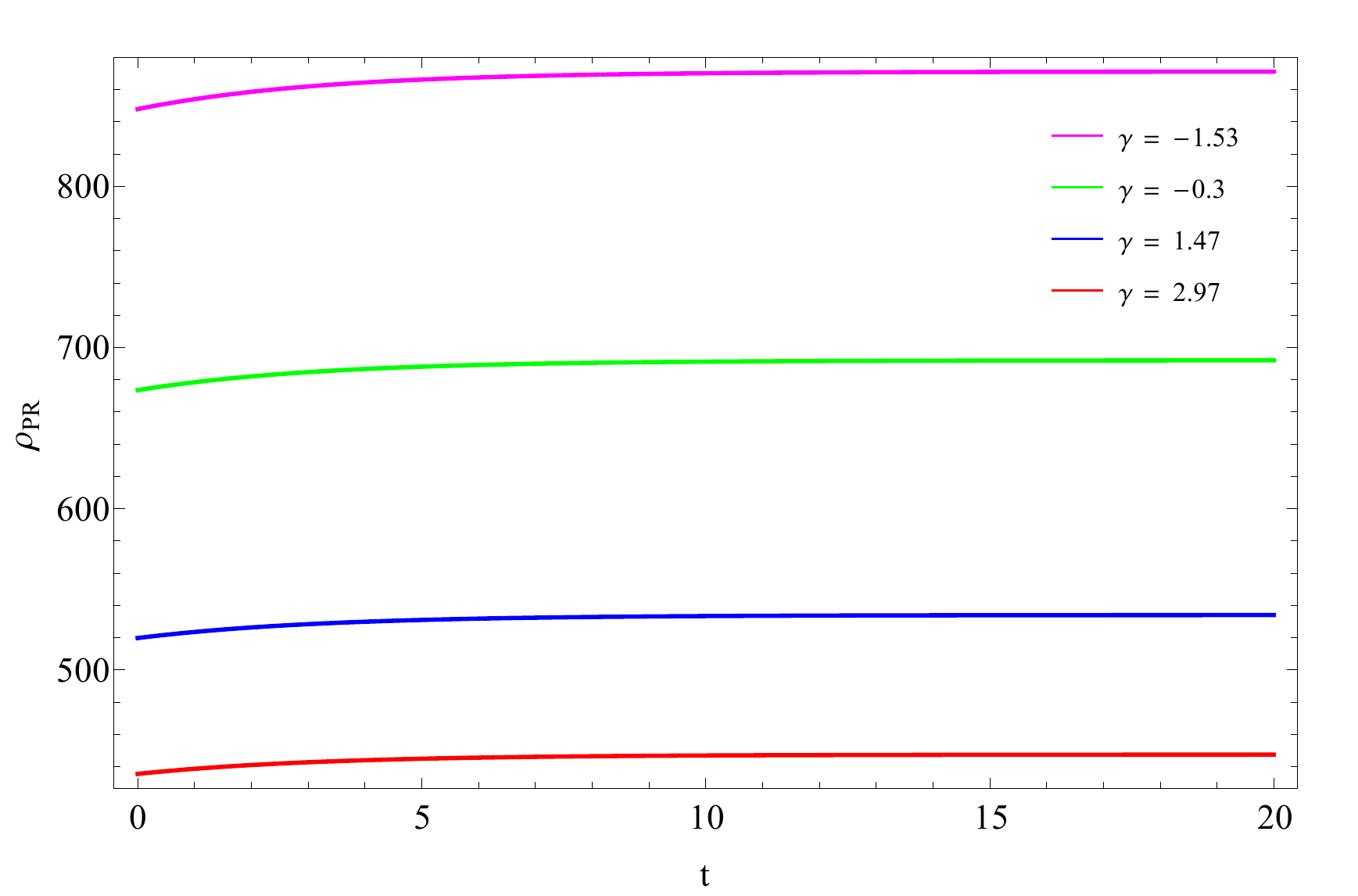}
\caption{Graphical behaviour of pressure (left panel) and energy density (right panel) of PR model in cosmic time  with the representative value of the parameter $\gamma$. The other parameters values are, $H_0=74.31$, $H_1=1$, $\lambda=0.3011$,  }
\label{Fig5}
\end{figure}

To analyse physically eqns. \eqref{eq:23} and \eqref{eq:24}, we need to fix two unknown parameters $\gamma$ and $\Lambda_{0}$. The value of cosmological constant at present epoch has been considered as $\Lambda_{0}= - 0.25$ from recent $f(R,T)$ framed models that have been widely studied  \cite{Tarai20}. But there is no observational evidence available for the value of $\gamma$, hence $\gamma$ can be adjusted precisely, $\gamma= -1.53,-0.03, 1.47, 2.97$ to obtain negative pressure and positive energy density. The same behaviour has been observed graphically for pressure Fig. \ref{Fig5} (left panel) and energy density Fig. \ref{Fig5} (right panel). $p_{PR}$ decreases and $\rho_{PR}$ increases over the time, but very slowly and at the late phase they almost remain constant.  These behaviours make our present model different from sudden future singularities in the sense that both $\mid p_{PR} \mid$ and $\rho_{PR}$ are finite and do not diverge \cite{Nojiri05}. \\

\begin{figure}[!htp]
\centering
\includegraphics[scale=0.50]{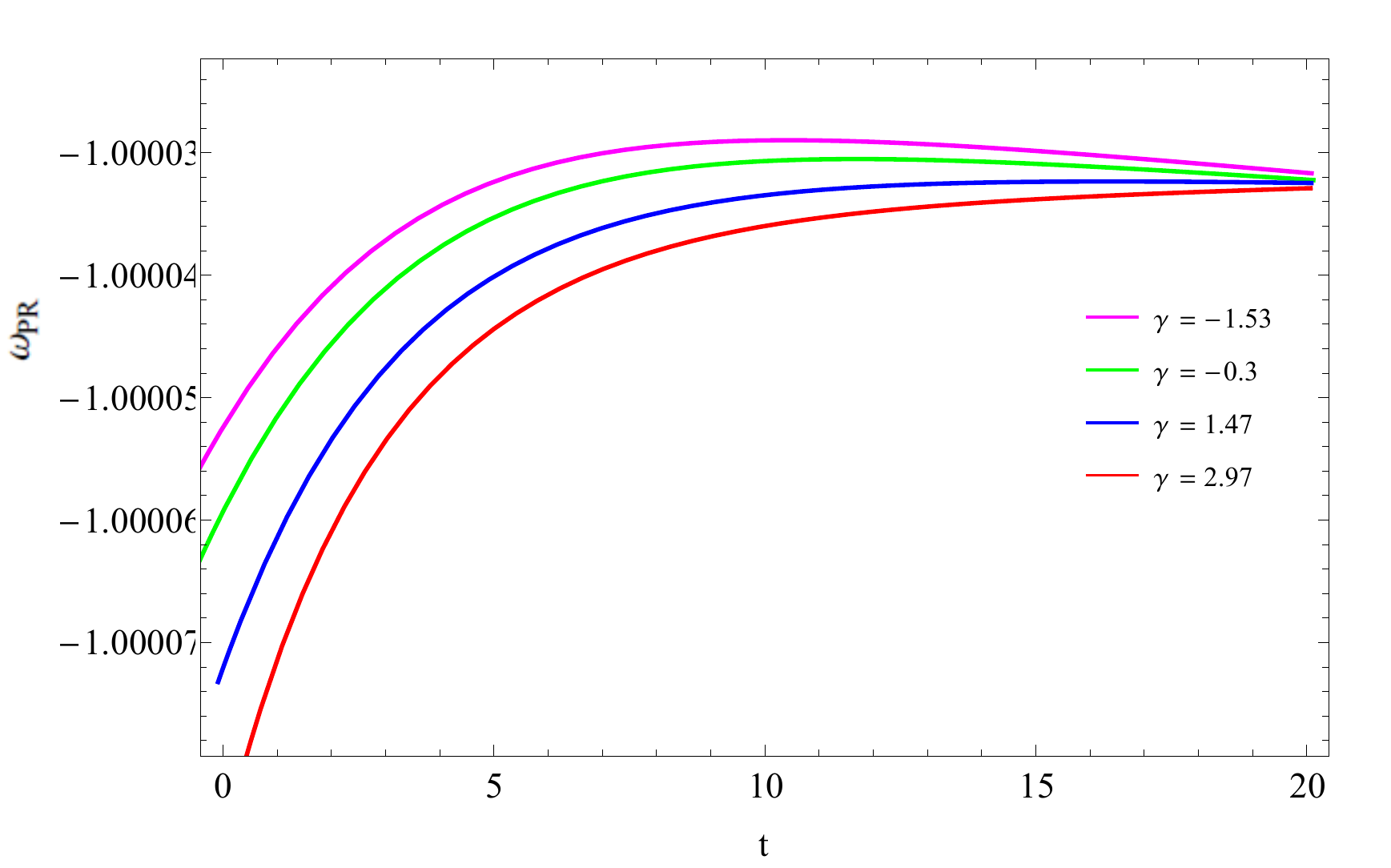}
\includegraphics[scale=0.50]{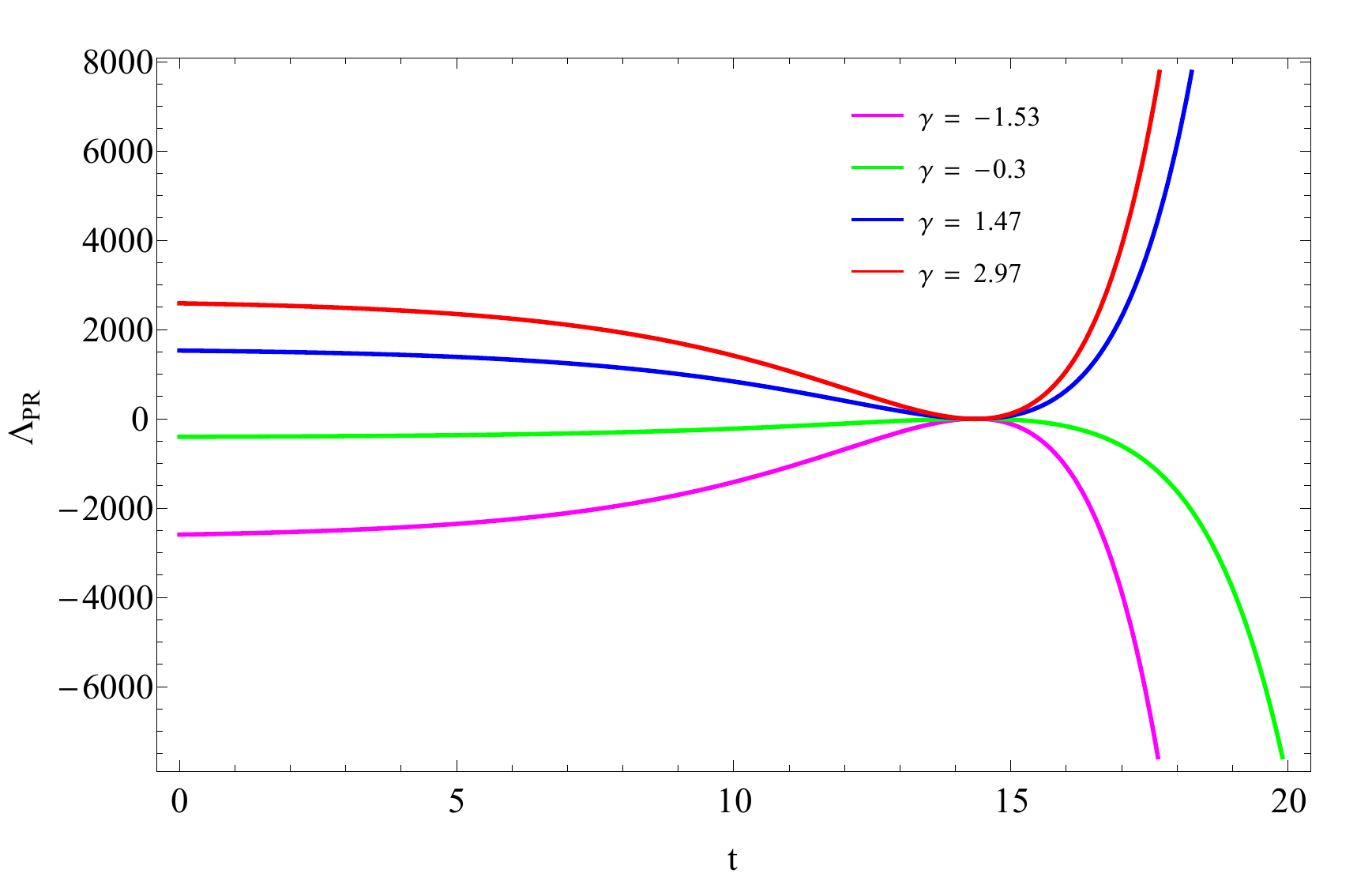}
\caption{Graphical behaviour of EoS parameter (left panel) and energy density (right panel) of PR model in cosmic time with the representative value of $\gamma$. The other parameters vales are, $H_0=74.31$, $H_1=1$, $k=1.0000814$, $\lambda=0.3011$, $\Lambda_0=-0.25$} \label{Fig6}
\end{figure}

Now, the EoS parameter of the PR model can be derived as,  
\begin{eqnarray}
\omega_{PR} &=& -1 \nonumber \\
&+&(\beta+\gamma) \left[\frac{3 f_{1}(k) f_{2}(k)\lambda H_{1} e^{-\lambda t}+9 f_{3}(k)(H_0-H_1e^{-\lambda t})^2}{\gamma\left[6 f_{2}(k)\lambda H_1e^{-\lambda t}+27(H_0-H_1e^{-\lambda t})^2\right]-\beta\left[9 f_{4}(k)(H_0-H_1e^{\lambda t})^2\right]+ f_{2}^{2}(k)(\beta-\gamma)\Lambda_0}\right], \label{eq:25}
\end{eqnarray}
where, $f_{1}(k)= k+1,$ $f_{2}(k)= k+2,$ $f_{3}(k)= k^{2}-k$ and $f_{4}(k)= 2k+1$. At early phase, $t\rightarrow 0$, 

\begin{equation}
\omega_{PR} = -1 +(\beta+\gamma)  \left[\frac{3 f_{1}(k) f_{2}(k)\lambda H_{1} +9 f_{3}(k)(H_0-H_1)^2}{\gamma\left[6 f_{2}(k)\lambda H_1 +27(H_0-H_1)^2\right]-\beta\left[9 f_{4}(k)(H_0-H_1)^2\right]+ f_{2}^{2}(k)(\beta-\gamma)\Lambda_0}\right] \nonumber
\end{equation}
and at late phase, $(t\rightarrow \infty)$, 
\begin{equation}
\omega_{PR} = -1 +(\beta+\gamma)\left[\frac{9 f_{3}(k)H_{0}^{2}}{27 H_{0}^{2}- 9 \beta f_{4}(k)H_{0}^{2}+ f_{2}^{2}(k)(\beta-\gamma)\Lambda_0}\right] \nonumber
\end{equation}

The dynamics of EoS parameter in an isotropic universe can derived by substituting $k=1$ in eqn. \eqref{eq:19}. This results in elimination of $f_{3}(k)$ and hence $\omega_{PR}$ asymptotically approaches to $-1$ at late epoch. Fig. \ref{Fig6} shows the evolutionary behavior of EoS parameter, which indicates that the PR model may overlap with $\Lambda$CDM model towards late phase of evolution. Also, the value of EoS parameter at present epoch as, $\omega_{PR}(t_{0})= -1.000015$. As value of $\gamma$ decreases  the rate of growth of $\omega_{PR}$ increases in the phantom region. Eventually, these $\omega_{PR}$ curves merge to depict the asymptotic value of  $\omega_{PR}$ as $-1$ towards late epoch. Now, the effective cosmological constant can be derived as,
\begin{eqnarray}
\Lambda_{PR} &=&\frac{\gamma}{(\beta+\gamma)}\left[\frac{6\lambda H_1 e^{-\lambda t}+18(H_0-H_1e^{-\lambda t})^2}{f_{2}(k)}-2\Lambda_0\right]+\Lambda_0\label{eq:26}
\end{eqnarray}

It is clear that, $\Lambda_{PR}$ depends on all model parameters discussed above, among which only $\gamma$ is the free parameter. If $\gamma$ vanishes, then the corresponding $f(R,T)$ framework for the model will be reduced to GR and $\Lambda_{PR}$ will be eventually same as the value of cosmological constant at present time $(\Lambda_{0}).$ Moreover, incorporating all fixed values of model parameters along with four different values of $\gamma$ at present cosmic time $(t\simeq 13.82 Gyr),$ the PR model predicts effective cosmological constant value around -0.25. This has been illustrated in Fig. \ref{Fig6}, which also indicates the evolution of effective cosmological constant depending on different coupling parameter values. Higher is the value of $\gamma$, higher is the growth rate of $\Lambda_{PR}$ curve shifting from negative to positive domain. For negative values of $\gamma,$ $\Lambda_{EF}$ turns out to be a negative quantity which increases with cosmic time in the early epoch and then decreases faster after attaining a maximum value at $t \simeq 14.25 $ Gyr. On the other hand, $\Lambda_{PR}$ becomes positive for positive $\gamma$ values and decreases initially in the early epoch. After attaining a minimum value at $t  \simeq 14.25 Gyr,$ it increases sharply towards late phase of evolution. Eventually, the curves of effective cosmological constant merge with growth of cosmic time  and asymptotically approaches to zero at near future at $t\simeq$ 14.25 Gyr. The reason of sudden deflection of $\Lambda_{PR}$ in the late epoch may be due to the exponential drive of dark energy in the late phase of cosmic evolution. Hence, it can be concluded that the effective cosmological constant is going to vanish in near future and then rapidly grow bearing some burden contributing towards an accelerating universe at late times.
  
\section{Energy conditions of the models}

The real universe consists of energy momentum tensor, which will create different matter fields and in turn knowing the precise form of matter field, it is not possible to identify the exact energy momentum tensor. There is limited idea on the behaviour of matter from the pressure and energy density. Therefore, it has been difficult to predict the occurrence of singularities in the universe from Einstein's field equations. However, there are certain inequalities, which is physically viable, called the energy conditions can address this issue. The energy conditions are, (i)  $T_{ij} n^{i}n^{j}\geq 0$, where $n^{i}$ is a null vector implying $\rho + p \geq 0$, Null Energy Condition (NEC); (ii) $T_{ij} v^{i}v^{j}\geq 0$, where $v^{i}$ is a time-like vector implying $\rho + p \geq 0,$ $\rho \geq 0$ , Weak Energy Condition (WEC), (iii) $T_{ij} v^{i}v^{j}\geq 0$, where $T_{ij}v^{i}$ is not space-like implying $\mid p \mid \leq \rho,$ $\rho \geq 0$, Dominant Energy Condition (DEC); and (iv) $T_{ij} v^{i}v^{j}\geq 0$, where $v^{i}$ is co-moving velocity vector implying $\rho + 3p \geq 0$, Strong Energy Condition (SEC). Violation of these energy conditions trigger both classical and quantum instabilities \cite{Carroll03}. As a result, many cosmological models admitting $\omega \leq -1$ suffer from severe disabilities in order to check the physical viability. In this view, few ideas have been explored so that the instabilities can be avoided \cite{Csaki05}. However, in spite of the violation of all energy conditions, phantom allows the BR singularity suitably. The energy conditions for BR model are, 
\begin{eqnarray}
\rho_{BR}+p_{BR}&=&\frac{-\beta}{\beta^{2}-\gamma^{2}}\left[ \frac{9\alpha^{2}(k^{2}-k)+3\alpha(k^{2}+3k+2)}{(k+2)^{2}(t_{s}-t)^{2}}\right] \nonumber\\
&+& \frac{\gamma}{\beta^{2}-\gamma^{2}}\left[ \frac{3\alpha(1+\alpha)(k^{2}-k-6)}{(k+2)^{2}(t_{s}-t)^{2}}\right]\nonumber\\
\rho_{BR}+p_{BR}&=&\frac{-\beta}{\beta^{2}-\gamma^{2}}\left[ \frac{9\alpha^{2}(k^{2}-k)+3\alpha(k^{2}+3k+2)}{(k+2)^{2}(t_{s}-t)^{2}}\right] \nonumber\\
&+& \frac{\gamma}{\beta^{2}-\gamma^{2}}\left[ \frac{3\alpha(1+\alpha)(k^{2}-k-6)}{(k+2)^{2}(t_{s}-t)^{2}}\right], \rho\geq0\nonumber\\
\rho_{BR}-p_{BR}&=&\frac{\beta}{\beta^{2}-\gamma^{2}}\left[ \frac{3\alpha(1+2\alpha)(k^{2}+3k+2)}{(k+2)^{2}(t_{s}-t)^{2}}\right] \nonumber\\
&-& \frac{\gamma}{\beta^{2}-\gamma^{2}}\left[ \frac{9\alpha^{2}(k^{2}-k)+3\alpha(k^{2}+3k+2)}{(k+2)^{2}(t_{s}-t)^{2}}\right]-\frac{2\Lambda_{0}}{\beta+\gamma}\nonumber\\
\rho_{BR}+3p_{BR}&=&\frac{\beta}{\beta^{2}-\gamma^{2}}\left[ \frac{9\alpha^{2}(-3k^{2}-k-2)-9\alpha(k^{2}+3k+2)}{(k+2)^{2}(t_{s}-t)^{2}}\right] \nonumber\\
&+& \frac{\gamma}{\beta^{2}-\gamma^{2}}\left[ \frac{3\alpha(3k^{2}+k-10)+27\alpha^{2}(k^{2}-k-42)}{(k+2)^{2}(t_{s}-t)^{2}}\right]+\frac{2\Lambda_{0}}{\beta+\gamma}\label{eq:27}
\end{eqnarray}

and for PR model,

\begin{eqnarray}
\rho_{PR}+p_{PR}&=&\frac{-\beta}{\beta^{2}-\gamma^{2}}\left[ \frac{3(k^{2}+3k+2)\lambda H_{1}e^{-\lambda t}+9(k^{2}-k)(H_{0}-H_{1}e^{-\lambda t})^{2}}{(k+2)^{2}}\right] \nonumber\\
&-& \frac{\gamma}{\beta^{2}-\gamma^{2}}\left[ \frac{3(k^{2}+k-2)\lambda H_{1}e^{-\lambda t}+9(k^{2}-k-3)(H_{0}-H_{1}e^{-\lambda t})^{2}}{(k+2)^{2}}\right]\nonumber\\&+&\frac{\Lambda_{0}}{\beta+\gamma}\nonumber\\
\rho_{PR}+p_{PR}&=&\frac{-\beta}{\beta^{2}-\gamma^{2}}\left[ \frac{3(k^{2}+3k+2)\lambda H_{1}e^{-\lambda t}+9(k^{2}-k)(H_{0}-H_{1}e^{-\lambda t})^{2}}{(k+2)^{2}}\right] \nonumber\\
&-& \frac{\gamma}{\beta^{2}-\gamma^{2}}\left[ \frac{3(k^{2}+k-2)\lambda H_{1}e^{-\lambda t}+9(k^{2}-k-3)(H_{0}-H_{1}e^{-\lambda t})^{2}}{(k+2)^{2}}\right]\nonumber\\&+&\frac{\Lambda_{0}}{\beta+\gamma}, \rho\geq0 \nonumber\\
\rho_{PR}-p_{PR}&=&\frac{\beta}{\beta^{2}-\gamma^{2}}\left[ \frac{3(k^{2}+k+2)\lambda H_{1}e^{-\lambda t}+9(k^{2}+3k+2)(H_{0}-H_{1}e^{-\lambda t})^{2}}{(k+2)^{2}}\right] \nonumber\\
&+& \frac{\gamma}{\beta^{2}-\gamma^{2}}\left[ \frac{3(k^{2}+k-2)\lambda H_{1}e^{-\lambda t}+9(k^{2}-k-3)(H_{0}-H_{1}e^{-\lambda t})^{2}}{(k+2)^{2}}\right]\nonumber\\&-&\frac{\Lambda_{0}}{\beta+\gamma}\nonumber\\
\rho_{PR}+3p_{PR}&=&\frac{-\beta}{\beta^{2}-\gamma^{2}}\left[ \frac{9(k^{2}+3k+2)\lambda H_{1}e^{-\lambda t}+9(3k^{2}+k+2)(H_{0}-H_{1}e^{-\lambda t})^{2}}{(k+2)^{2}}\right] \nonumber\\
&-& \frac{3\gamma}{\beta^{2}-\gamma^{2}}\left[ \frac{3(k^{2}+k-2)\lambda H_{1}e^{-\lambda t}+9(k^{2}-k-3)(H_{0}-H_{1}e^{-\lambda t})^{2}}{(k+2)^{2}}\right]\nonumber\\&+&\frac{3\Lambda_{0}}{\beta+\gamma} \label{eq:28}
\end{eqnarray}

\begin{figure}[!htp]
\centering
\includegraphics[scale=0.50]{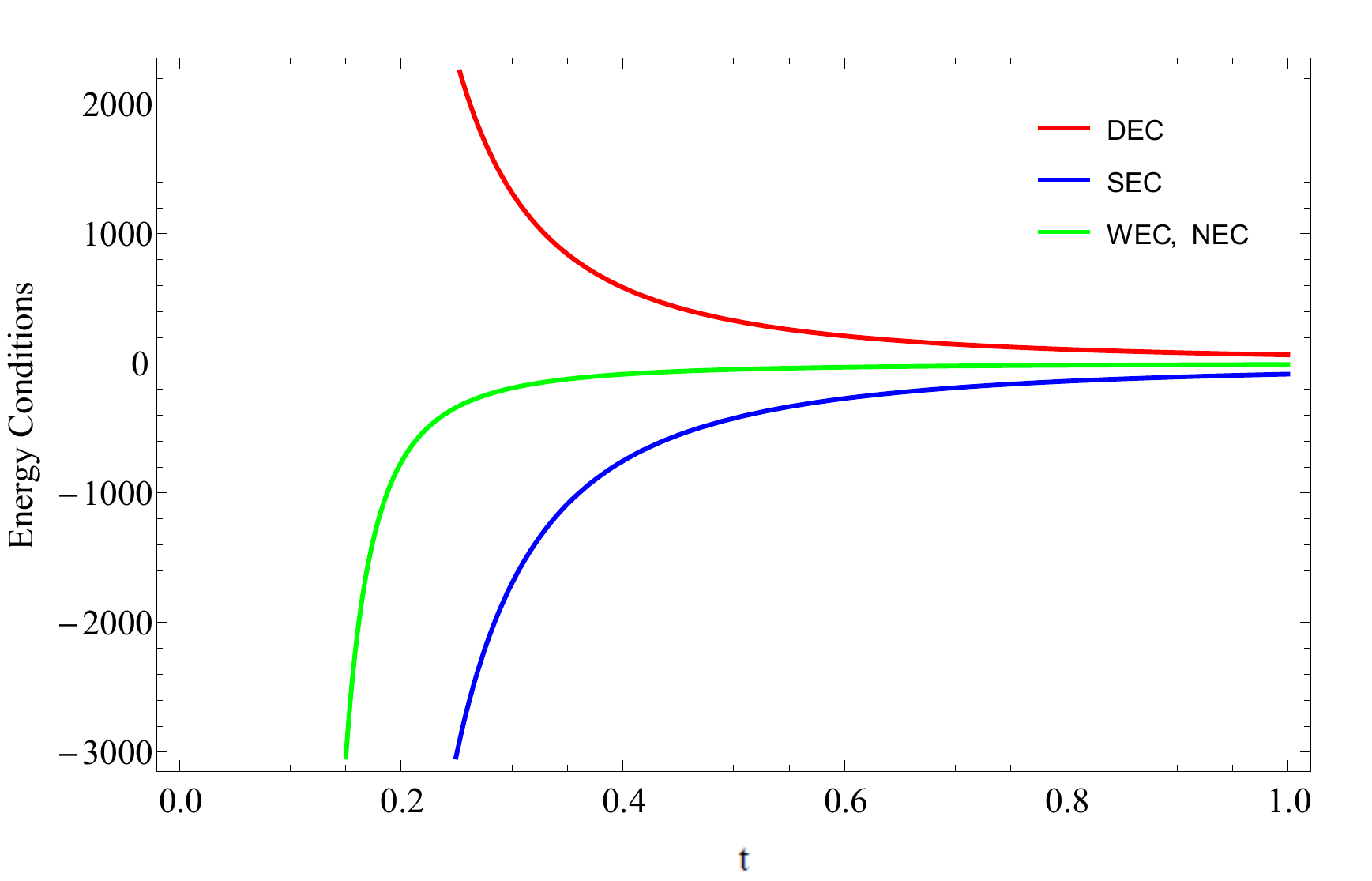}
\includegraphics[scale=0.50]{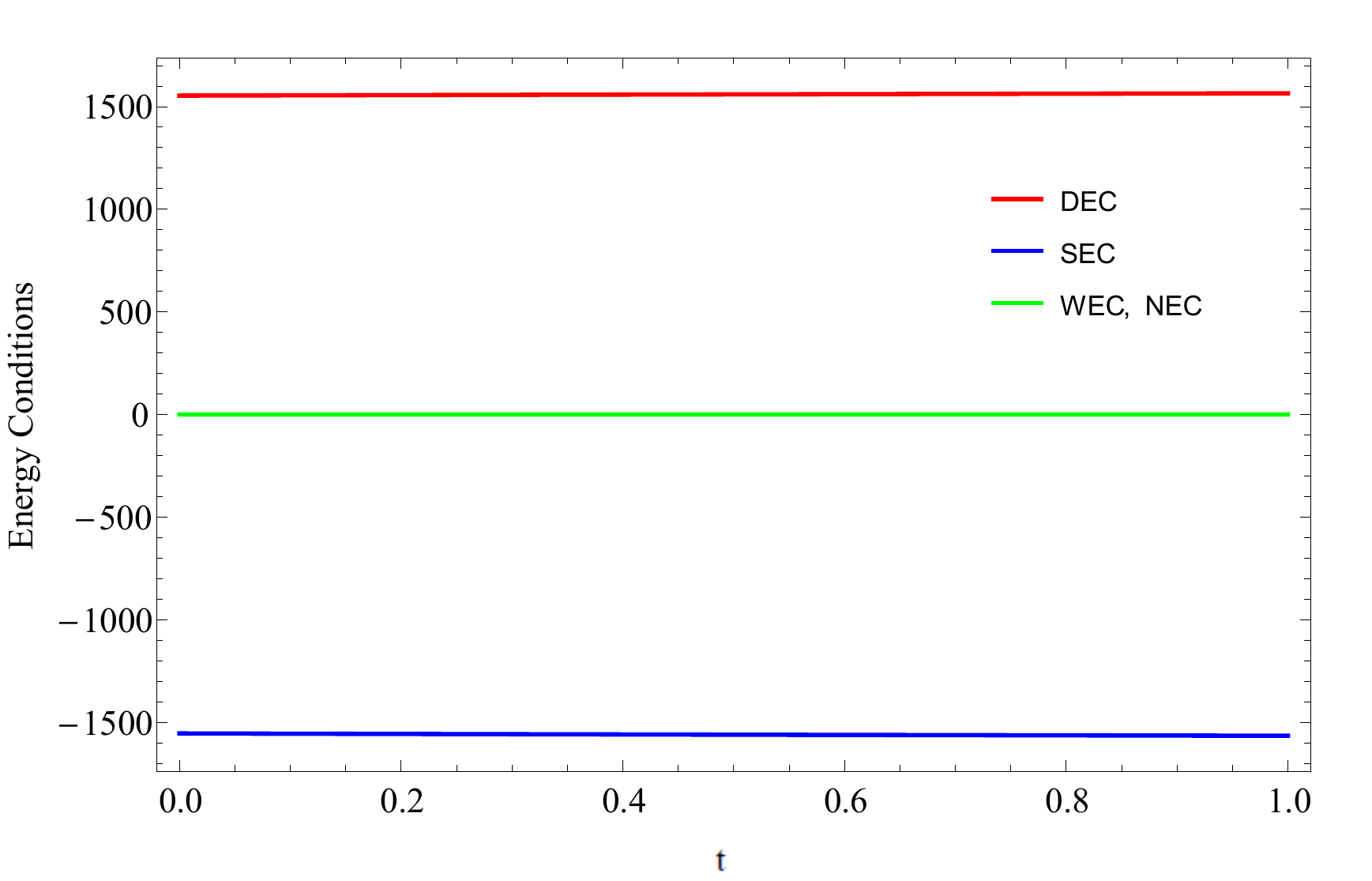}
\caption{Graphical behaviour of energy conditions for BR model (left panel) and energy PR model (right panel)in cosmic time. The parameters values for BR graph, $k=1.0000814$, $\alpha=12.5$, $t_s=0.1$, $\Lambda_0=-0.25$ and for PR graph, $H_0=74.31$, $H_1=1$, $k=1.0000814$, $\lambda=0.3011$, $\Lambda_0=-0.25$, }
\label{Fig7}
\end{figure}

Since both the models evolve in the phantom region as discussed earlier, the energy conditions are expected to be violated except the DEC. Fig. \ref{Fig7}(left panel) and Fig \ref{Fig7} (right panel) graphically represent the behaviour of energy conditions for BR and PR model respectively. For both the models, only DEC is satisfied with in proper range but WEC and SEC are violated as expected. However, for BR model, the energy condition curves seems to be merged along with growth of cosmic time, whereas in case of PR model, the curves are constant throughout.

\section{Geometrical behaviour of the model and conclusions}
In addition to the geometrical parameters $H$ and $q$, there are other geometrical parameters such as the jerk parameter $j$ and snap parameter $s$, which are respectively used the third and fourth derivative of the scale factor. The $(j,s)$ pair called the state finder diagnostic pair helps in differentiating different dark energy models \cite{Sahni03}. To mention, we must consider the fact that BR and PR singularities are related to a sudden blow-up of the corresponding Hubble parameter, scale factors and their respective time derivatives. Now, these two diagnostic parameters can be obtained as, 

\begin{equation}
j= \frac{1}{a H^{3}} \left[\frac{{d}^3 a(t)}{dt^{3}} \right]
\end{equation}
and
\begin{equation}
s= - \dfrac{1}{a H^{4}} \left[\frac{{d}^4 a(t)}{dt^{4}}  \right]
\end{equation}

\begin{figure}[h!]
\includegraphics[width=45mm]{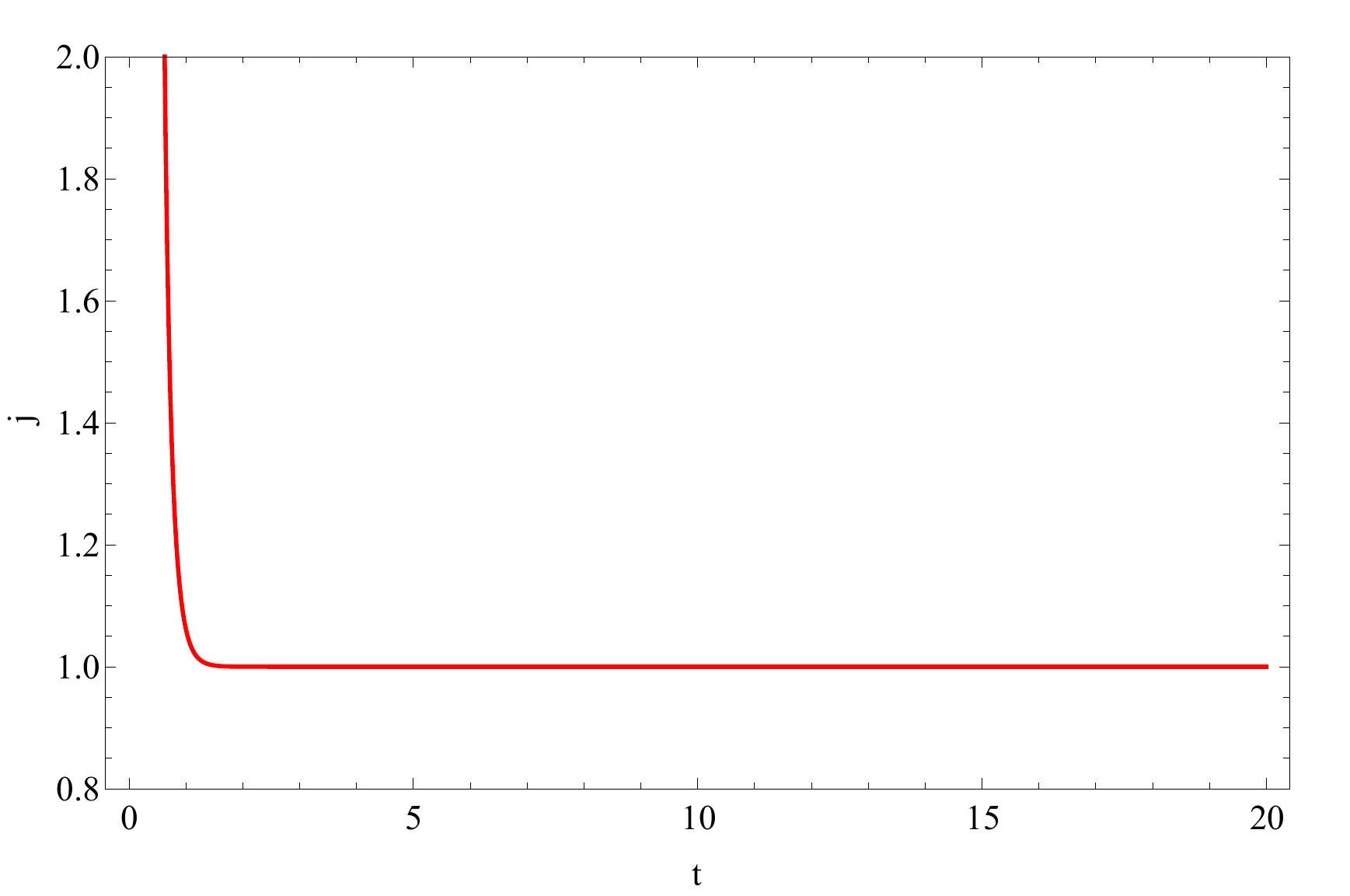}
\includegraphics[width=45mm]{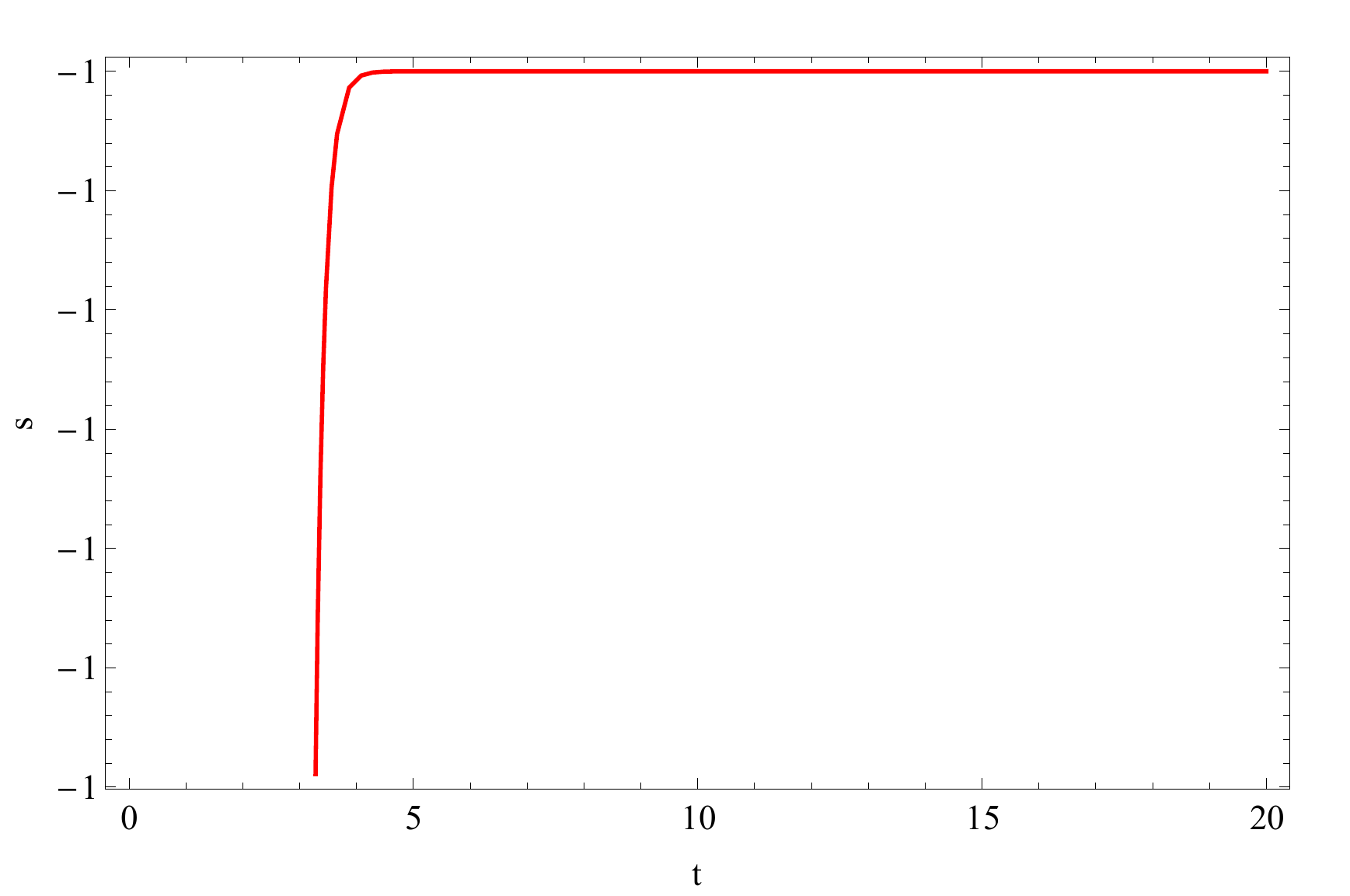}
\includegraphics[width=45mm]{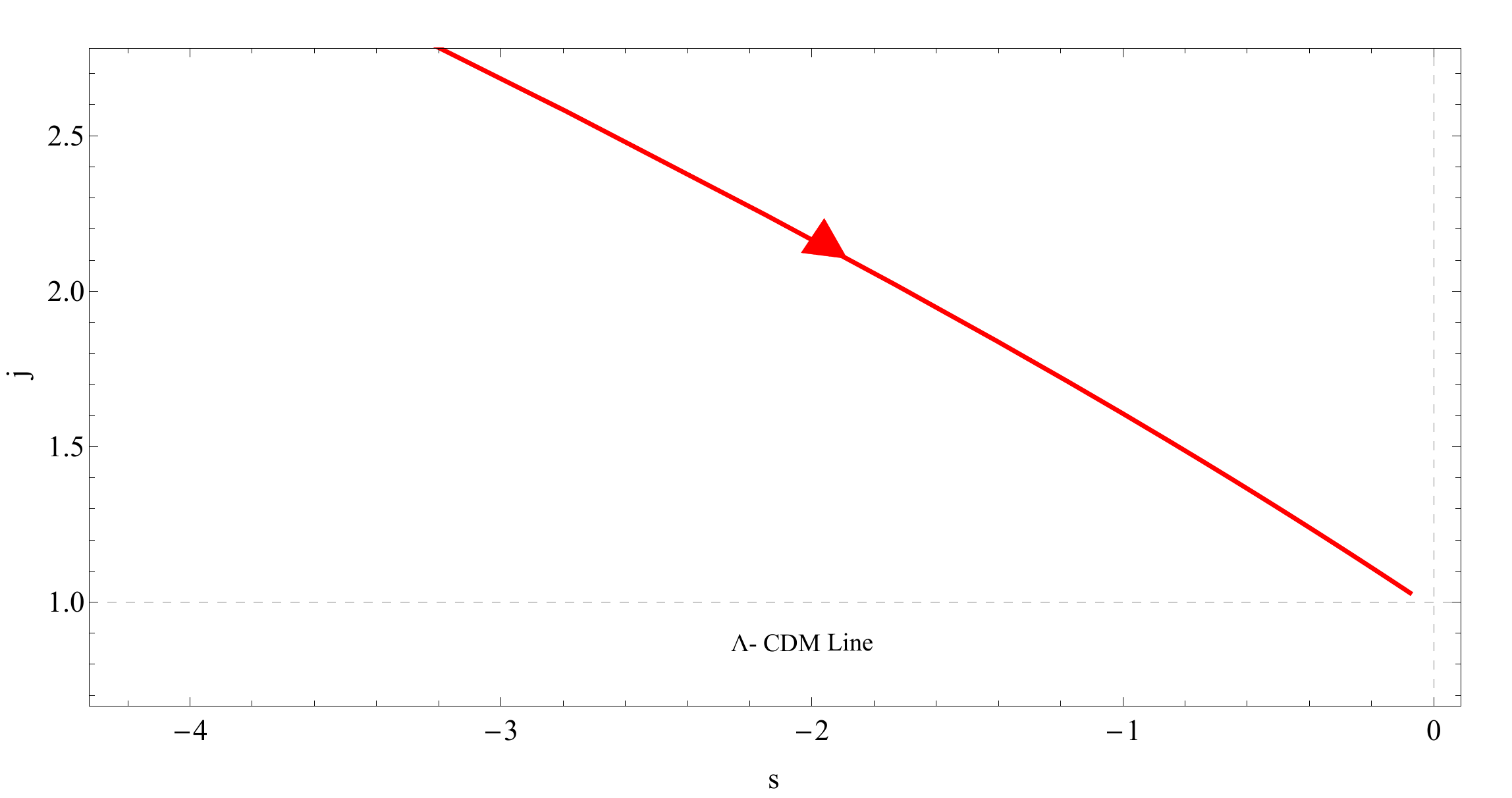}
\caption{Evolution of jerk parameter (left panel), snap parameter(middle pane) in cosmic time and $j-s$ behaviour(right panel).}\label{Fig8}
\end{figure}

The corresponding time derivatives can be expressed by a general iterative formula as;

\begin{equation}
X^{(n)} = H [n (q+1) X^{(n)}- (X^{(n+1)}+ X^{(n)})],
\end{equation}
where $n$ denotes the order of the derivative. The state finder pair $(j,s)$ can predict an emergence of BR and PR singularities, giving suitable signals to generalized sudden future singularities. In fact, the signal for BR singularity is mentioned in the form, $\mid H_{0} \mid \rightarrow \infty,$ for \textit{sudden future singularity},  $\mid q_{0} \mid \rightarrow \infty$ and for \textit{generalized sudden future singularity},  $\mid j_0 \mid \rightarrow \infty,$  $\mid s_0 \mid \rightarrow \infty$ \cite{Caldwell02,Darbowski06}. Hence, it is clear that the state finder pair can provide more important and sensitive analysis to the current status of matter. Moreover, a recent observational investigation on type Ia Supernova data \cite{Riess18} confirms the domain of the state finder pair on $j-s$ plane as, $j(t_0) > 0$ and $s(t_0) \leq 0$. The same features have been shown in Fig. \ref{Fig8}. The evolution of state finder pair which approaches to the point $(1,0)$ towards late epoch in order to overlap with $\Lambda$CDM model Fig. \ref{Fig8} (right panel).

Finally, we have presented the BR and PR cosmological models of the universe at the backdrop of an anisotropic universe and by extending the geometrical part of Einstein-Hilbert action. The present value of Hubble parameter has been obtained as $74.33$ for BR model and  $74.31$ for PR model by fixing the deceleration parameter at $-1.08$. Both of these $H$ value correspond to the observational results and hence we can confirm the viability of the scale factor considered to frame the cosmological models. Theoretically the value of  EoS parameter has been significant to address the evolution history of the universe. In the BR model, for varying $\beta$ the evolution of the universe is very slow and with the higher $\beta$ value, the behaviour of the model gradually leading to phantom behaviour. At the same time, in varying moment the EoS evolves from a lower negative value to higher value and there is a sudden change in the behaviour after some time and further decreases. At the late time it mostly remains in phantom region and the reason might be the dominance of phantom phase on the evolution process. Further study on this claims that the BR scenario would have occurred before $22Gyr$ before the present time. Interestingly in PR model, the evolution starts from higher negative value and at present time, it attains $\omega_{PR}(t_0)\thickapprox-1.000015$, which is in the range of observational findings. Further at late time it may overlap with the $\Lambda$CDM line. Another feature of the model is the behaviour of energy conditions, where in the context of extended gravity, the SEC should violate. In BR model,the SEC violates and the NEC appear to be merged along the cosmic time. In PR, model, the violation of SEC and merging of NEC with the time axis has been observed. An interesting feature is that the energy conditions remains same throughout the evolution. Since both the models are favouring phantom behaviour, it is expected that both SEC and NEC should violate, however the merging of NEC is also acceptable with some restrictions. Nevertheless, both models are showing appropriate behaviour that based on the cosmological observations. 

\section*{Acknowledgement}
ST acknowledges Rashtriya Uchachatar Shikshya Abhiyan(RUSA), Ministry of HRD, Govt. of India for the financial support. BM and SKT acknowledges Inter-University Center for Astronomy and Astrophysics (IUCAA), Pune, India for hospitality and support during an academic visit where a part of this work is accomplished.

\end{document}